\documentclass[
aps,
prfluids,
onecolumn,
superscriptaddress,
nofootinbib,
floatfix,
longbibliography,
]{revtex4-2}
\usepackage{amsmath,amssymb,bm}
\usepackage{graphicx}
\usepackage{dcolumn}
\usepackage{xcolor}
\usepackage{hyperref}
\usepackage{subcaption}

\usepackage{float}
\usepackage{placeins}
\usepackage{bigints}
\usepackage{lineno}

\graphicspath{{./Figs/}}

\begin{document}
	
	\title{Solutocapillary instability in slipping falling films}
	
	\author{Sanghasri Mukhopadhyay}
	\email{sanghasri.m@iiitb.ac.in}
	\affiliation{Centre for Applied Sciences (CAS), International Institute of Information Technology, Bengaluru 560100, India}
	
	\author{S\'everine Millet}
	\affiliation{Universit\'e Claude Bernard Lyon 1, LMFA, UMR 5509, CNRS,
		Ecole Centrale de Lyon, INSA Lyon, 69622 Villeurbanne, France}
	
	\author{Bastien Di Pierro}
	\affiliation{Universit\'e Claude Bernard Lyon 1, LMFA, UMR 5509, CNRS,
		Ecole Centrale de Lyon, INSA Lyon, 69622 Villeurbanne, France}
	
	\author{Asim Mukhopadhyay}
	\affiliation{Department of Mathematics, Vivekananda Mahavidyalaya,
		Burdwan 713103, West Bengal, India}
	
	\date{\today}
	
	\maketitle
		We present a comprehensive framework for gravity-driven, surfactant-laden thin films flowing over slippery substrates, elucidating how wall
		slip modifies the coupled hydrodynamics and interfacial transport. A
		long-wave model is formulated with a conservative bulk–surface mass
		balance and a Navier slip condition. The Orr–Sommerfeld eigenvalue problem governs the linear regime, while a weighted-residual model captures the nonlinear evolution over a range of equilibrium surfactant coverages, Marangoni strengths, and adsorption kinetics. The analysis predicts a non-monotonic variation of the critical Reynolds number
		with equilibrium coverage, exhibiting a maximum at intermediate $\Gamma_e$, and a slip-induced transition from single- to double-hump solitary structures with increasing Marangoni number, accompanied by attenuated capillary ripples. Under fast adsorption kinetics, the surface field homogenizes, preserving the mean film shape and flux while flattening both the surface concentration $\Gamma$ and the bulk inventory $\chi + h\phi$. A spurious interfacial mass
		growth reported by \citet{pascalStabilityInclinedFlow2019} and \citet{dalessioMarangoniInstabilitiesAssociated2020} is
		resolved through a revised surface balance ensuring strict conservation. Wall slip thus emerges as a key control parameter, reducing viscous
		resistance and mitigating Marangoni back-stress. The slip parameter $\beta$ is a useful control knob for surfactant-laden films.  Slip prevents fragile multi-hump bound states, promoting a single broad crest or an almost flat, uniform sheet by carefully bonding $\beta$ to wave selection, ripple damping, and the bulk-surface surfactant balance.
		


	\section{\label{sec:introduction}Introduction}
	
	Gravity-driven liquid films flowing down inclined surfaces occur in many natural and industrial settings,
	including coating flows, heat and mass transfer devices, and absorption systems.
	Such films are known to become unstable beyond a critical flow rate, leading to the formation of interfacial
	waves whose onset and nonlinear evolution have been studied extensively.
	
	Surfactants, or surface-active agents, preferentially adsorb at fluid interfaces and modify surface tension,
	thereby influencing interfacial dynamics.
	Their role in fluid systems is well documented in both industrial and biological contexts, ranging from
	absorption refrigeration systems to pulmonary mechanics.
	Early theoretical studies by \citet{whitakerEffectSurfaceActive1964} and \citet{benjaminEffectsSurfaceContamination1964}
	demonstrated that insoluble surfactants can have a pronounced stabilizing effect on the primary instability
	of falling films through Marangoni stresses, effectively imparting an elastic character to the interface.
	
	Subsequent work investigated the nonlinear evolution and rupture of thin surfactant-laden films.
	For example, \citet{jensenInsolubleSurfactantSpreading1992} showed that surfactant-induced capillary forces can drive
	film thinning and rupture, while nonlinear analyses by \citet{dewitNonlinearEvolutionEquations1994} and
	\citet{desouzaPatternFormationThin1998} identified conditions for pattern formation.
	Over the past two decades, numerous studies have examined the influence of insoluble surfactants on falling-film
	instabilities under the effects of gravity, heating, and inertia
	\citep{blythEffectSurfactantStability2004, levyGravitydrivenThinLiquid2007, anjalaiahThinFilmFlow2013,
		hermansLungSurfactantsDifferent2015, bhamlaInterfacialMechanismsStability2017, srivastavaEffectInsolubleSurfactant2018,
		huFallingFilmInsoluble2020}.
	Using the weighted residual integral boundary layer method, \citet{pereiraDynamicsFallingFilm2008} analyzed the
	dynamics of surfactant-laden films on inclined substrates.
	
	In practice, however, most surfactants are at least partially soluble, allowing exchange between the interface
	and the bulk liquid.
	This exchange alters interfacial concentration gradients and can weaken or suppress Marangoni stresses when
	transfer rates are high.
	Experimental studies by \citet{georgantakiEffectSolubleSurfactants2012} and \citet{katsiavriaStabilityLiquidFilm2020}
	demonstrated that increasing surfactant solubility leads to marked changes in film dynamics and stability.
	
	Theoretical studies have likewise shown that soluble surfactants introduce new instability mechanisms.
	\citet{jiEffectHeatTransfer1995} identified Marangoni-driven instabilities in falling films contaminated with
	soluble surfactants, while \citet{shkadovFallingFilmsMarangoni2004} reported multiple diffusion-driven modes.
	Further analysis by \citet{yiantsiosMechanismMarangoniInstability2010} highlighted the competing roles of solubility,
	diffusion, and surface viscosity in evaporating films.
	More recently, \citet{karapetsasPrimaryInstabilityFalling2013} and \citet{karapetsasRoleSurfactantsMechanism2014}
	performed a systematic study of soluble-surfactant effects on inclined flows, showing that increased solubility
	can destabilize the film.
	These findings were supported experimentally by \citet{georgantakiMeasurementsStabilisationLiquid2016}, who observed
	a significant increase in the critical Reynolds number due to surfactant presence.
	
	The combined influence of surfactant transport and additional physical effects has also received attention.
	\citet{pascalStabilityInclinedFlow2019} examined films with variable mass density, revealing a non-monotonic
	dependence of stability on surfactant concentration.
	The interplay between solutal and thermocapillary effects was investigated by
	\citet{dalessioMarangoniInstabilitiesAssociated2020}, who showed that surfactants can enhance inertial instabilities
	in non-isothermal flows.
	In an isothermal setting, \citet{samantaRoleSolubleSurfactant2025} revisited a similar configuration and reported
	that surfactants may stabilize long-wave disturbances while destabilizing compliant-substrate flows.
	Related Orr-Sommerfeld analyses for non-isothermal films were presented by
	\citet{samantaEffectSolubleSurfactant2025}, who showed that unstable surfactant modes may disappear in the absence
	of external shear.
	
	In the classical description of falling films over solid substrates, the no-slip boundary condition is commonly
	assumed.
	However, the strict enforcement of no slip has been shown to lead to stress singularities at moving contact lines
	\citep{vMotionFluidfluidInterface1974, mukhopadhyayHydrodynamicInstabilityWave2021}, and experiments have reported
	finite slip on hydrophobic surfaces \citep{churaevSlippageLiquidsLyophobic1984}.
	A widely used alternative is the Navier slip boundary condition \citep{navierMemoireLoisMouvement1823}, which
	relates the wall shear stress to the tangential velocity.
	Its physical basis has been supported by molecular dynamics simulations
	\citep{delatorreMicroscopicSlipBoundary2019} and reviewed by \citet{netoBoundarySlipNewtonian2005}.
	Despite this, the role of wall slip in falling films laden with soluble surfactants has not been fully clarified.
	
	In addition to examining the hydrodynamic–surfactant coupling in the presence of wall slip, the present work
	addresses an inconsistency reported in previous long-wave models, where a non-physical accumulation of surfactant
	mass at the interface was observed during nonlinear evolution
	\citep{pascalStabilityInclinedFlow2019, dalessioMarangoniInstabilitiesAssociated2020}.
	Here, we investigate the stability and nonlinear dynamics of gravity-driven, surfactant-laden falling films over an inclined substrate with wall slip.
	We perform a linear stability analysis based on the Orr-Sommerfeld formulation to determine how the critical
	Reynolds number depends on the slip parameter and the base-state surfactant concentration.
	We then develop a depth-averaged weighted residual model that enforces a strictly conservative bulk-surface
	surfactant balance, allowing us to isolate the coupled effects of wall slip and surfactant solubility on the
	onset of instability and nonlinear wave dynamics.\par 
	While introducing slip modifies only one boundary condition, it changes the near-wall shear and therefore the
	base advection that drives surfactant redistribution and Marangoni back-stress. This coupling alters both the linear onset threshold and the nonlinear wave selection, which cannot be inferred from the no-slip theory by a simple parameter shift.

	The paper is organized as follows.
	Section~II presents the governing equations and nondimensionalization.
	Section~III describes the linear stability analysis in both the arbitrary-wavelength and long-wave limits.
	Section~IV examines nonlinear wave evolution, and Section~V summarizes the main conclusions.

	\section{\label{sec:formulation} Problem statement and formulation}
	
	\begin{figure}
		\centering
		\includegraphics[scale=.15]{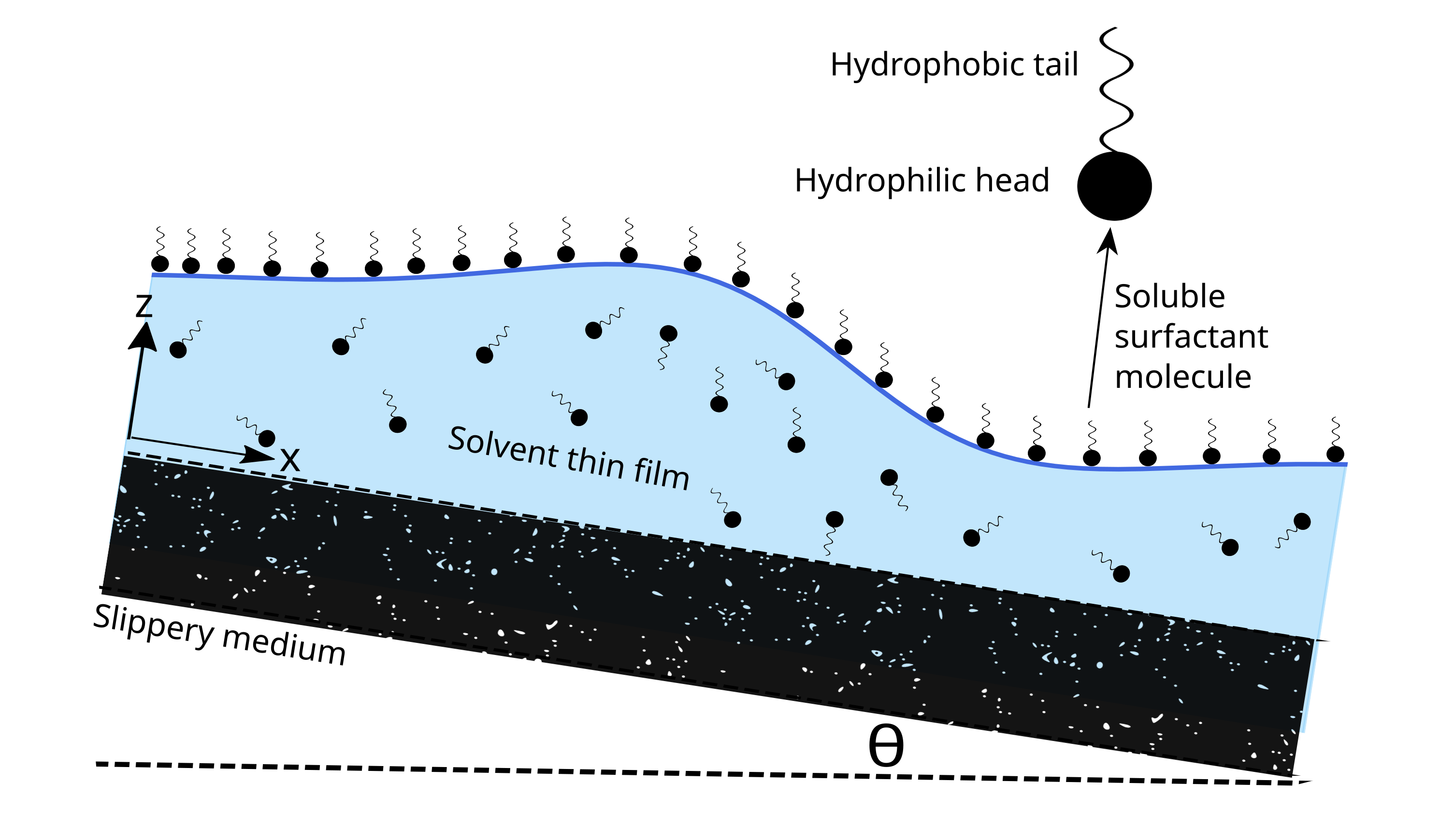}
		\caption{\label{fig:diagram}Sketch of the problem }
	\end{figure}
	We consider a thin Newtonian liquid film (viscosity $\eta$, density $\rho$) flowing down an inclined slippery substrate at angle $\theta$ to the horizontal.
	The free surface is located at $z=h(x,t)$.
	The flow is two-dimensional, gravity-driven, and the liquid contains a soluble surfactant.
	We suppose that a constant mass of surfactant is applied to the liquid layer.  Because we are working with a soluble surfactant, we must consider two different concentrations: one for the surfactant adsorbed at the surface and one for the surfactant dissolved in the bulk of the layer.  Furthermore, we will assume that the concentrations of surfactants are below the critical level for micelle formation, such that the surfactant added to the liquid exists as monomers.

	The coordinate system is chosen such that the $x-\text{axis}$ along the flow direction and the $z-\text{axis}$ normal to the inclined plane (see Figure \ref{fig:diagram}).\par
	
	The basic governing equations for the problem (\cite{dalessioMarangoniInstabilitiesAssociated2020}) are the equation of continuity:
	
	\begin{equation}
		\nabla \cdot \vec{V} = 0,
	\end{equation}
	the equation of momentum balance (Navier-Stokes equation for incompressible fluid):
	\begin{eqnarray}
		\rho D_t \vec{V}
		= - \nabla p + \rho \text{g} + \eta \nabla^2 \vec{V},
	\end{eqnarray}
	and the advection-diffusion equation for the surfactant:
	\begin{equation}
		D_t C = D_b\nabla^2 {C},
	\end{equation}
	$D_t = \partial_t + \vec{V} \cdot \nabla$ is the convective derivative, $\nabla$ denotes the two-dimensional gradient operator and $\nabla^2$ the Laplacian, $\vec{V} \equiv (u,w)$, where $u(t,x,z)$ and $w(t,x,z)$ denote the velocity components in the $x$ and $z$ directions, respectively, $t$ denotes the time, $p(t,x,z)$ denotes the pressure, $\textsl{g} = (g \sin\theta, -g\cos\theta)$ is the gravitational acceleration, $C(t,x,z)$ is the bulk concentration of the surfactant and $D_b$ is the diffusivity of the surfactant in the bulk fluid. \par 
	The pertinent boundary conditions on the plane substrate $z = 0$ (\cite{mukhopadhyayHydrodynamicInstabilityWave2021}) are
	
	\begin{equation}\label{boundary_bottom_dimension}
		\vec{V}\cdot \mathbf{t}_w = l_s (\mathbf{n}_w \cdot \tau \cdot \mathbf{t}_w), \quad \vec{V} \cdot \mathbf{n}_w = 0, \quad \mathbf{n}_w \cdot \nabla C = 0,
	\end{equation}
	where, $\mathbf{t}_w$ and $\mathbf{n}_w$ are unit tangential and unit normal vectors on the solid-liquid interface, respectively \citep{zhaoThermodynamicallyConsistentModel2021}, and $l_s$ is the (dimensional) Navier slip length. The dimensionless slip parameter used later is $\beta=l_s/\hat{h}_N$ \cite{mukhopadhyayHydrodynamicInstabilityWave2021}). \citet{voronovReviewFluidSlip2008} experimentally determined that an estimate of the slip length for a polydimethylsiloxane (PDMS) slippery plane is 250 $\mu$m \citep{voronovReviewFluidSlip2008}. The first equation of (\ref{boundary_bottom_dimension}) is called the Navier slip boundary condition, which states that the velocity at the boundary is proportional to the tangential component of the wall stress. The second equation of (\ref{boundary_bottom_dimension}) is the non-penetration condition, which implies that the fluid is not allowed to cross the boundary. The third equation of (\ref{boundary_bottom_dimension}) is the non-mass-flux condition of the solute.\par
	
	The boundary conditions on the free surface $z=h(x,t)$, separating the liquid layer from a quiescent ambient gas,
	are given by (\cite{mukhopadhyayWavesInstabilitiesViscoelastic2020})

	\begin{equation}\label{boundary_pressure_dimension}
		p_a + \mathbf{n}\cdot\tau\cdot\mathbf{n} = -\sigma \nabla\cdot\mathbf{n}
	\end{equation}
	\begin{equation}\label{boundary_stress_dimension}
		\mathbf{n} \cdot \tau \cdot \mathbf{t} = \nabla \sigma \cdot \mathbf{t}
	\end{equation}
	\begin{equation}\label{boundary_kinematic_dimension}
		D_tf = 0 
	\end{equation}
	
	are the usual normal and tangential stress balances along with the kinematic boundary condition, where $p_a$ is the pressure of the ambient gas, $\sigma$ is the surface tension coefficient of the fluid and $\mathbf{n}$, and $\mathbf{t}$ are the unit normal and tangential vectors on the free surface, respectively, and $f(x,t) = z-h(x,t)$, $\mathbf{\tau} = -p I + \eta (\nabla \vec{V} + (\nabla \vec{V})^T)$ is the total stress tensor, $I$ is the identity tensor. Equations (\ref{boundary_pressure_dimension}) and (\ref{boundary_stress_dimension}) are called dynamic conditions that arise from a balance between ambient pressure and surface tension. Equation (\ref{boundary_kinematic_dimension}) is a kinematic condition, which states that a particle on the free surface always remains there.\par
	
	Since we assumed that a fixed mass of soluble surfactant is added to the liquid layer, we must consider a separate concentration to describe the transport of surfactant absorbed at the interface. To formulate the surfactant transport equation at the free surface, we must consider advective and diffusive transport along the interface, as well as surfactant adsorption and desorption between the interface and the bulk solution. We perform the following equation \citep{gaverDynamicsLocalizedSurfactant1990, stoneSimpleDerivationTimedependent1990,manikantanSurfactantDynamicsHidden2020} to incorporate the above fact in $z = h(x,t)$

	\begin{equation}\label{surface_surfactant}
		\underbrace{\frac{\partial \Gamma}{\partial t}}_\text{unsteady} + \underbrace{\nabla_s \cdot(\Gamma \vec{V}_s)}_\text{advection}  = \underbrace{D_s \nabla_s^2 \Gamma}_\text{diffusion} + \underbrace{J_{bs}}_\text{source term},
	\end{equation}
	where, $\Gamma(x,t)$ is the concentration of surface surfactant, $\nabla_s = (I - \mathbf{nn})\cdot\nabla$ is the surface gradient operator, $\vec{V}_s= (I - \mathbf{nn})\cdot \vec{V}$ is the tangential velocity on the free surface. $D_s$ is the molecular diffusivity of the surface and $J_{bs}$ denotes the net flux of surfactant absorbed on the free surface. \par
	
	We want to emphasize there that $\nabla_s \cdot(\Gamma \vec{V}_s)$ is the surface advection term while $D_s \nabla_s^2 \Gamma$ is the surface diffusion term.  We would like to highlight that in this model the surfactant is permitted to diffuse along the interface with constant diffusivity or to be advected along the thin film's surface \citep{gaverDynamicsLocalizedSurfactant1990, stoneSimpleDerivationTimedependent1990, rahmanNanoscaleSurfactantTransport2025}. \par
	We also assumed that the adsorption rate is proportional to the concentration in the bulk and that the desorption rate is proportional to the surface concentration \citep{karapetsasRoleSurfactantsMechanism2014, dalessioMarangoniInstabilitiesAssociated2020}. Then according to Langmuir isotherm \citep{edwardsInterfacialTransportProcesses2013}, at $z = h(x,t)$
	
	\begin{equation} \label{Langmuir}
		J_{bs} = k_1\left(1 - \frac{\Gamma}{\Gamma_\infty}\right)C - k_2\Gamma,
	\end{equation}
	where, $k_1$ and $k_2$ are the adsorption and desorption reaction rates, respectively, and $\Gamma_\infty$ denotes the maximum packing surface concentration. For insoluble surfactant, the source term $J_{bs}=0$, as there is no adsorption/desorption, so that $k_1$ and $k_2$ are essentially zero \citep{palmerHydrodynamicStabilitySurfactant1972}. \par
	According to Fick's law of diffusion, the molar flux resulting from diffusion is proportional to the bulk concentration gradient of the surfactant; we have the following
	\begin{equation}
		J_{bs} = - D_b \nabla C \cdot \mathbf{n}.
	\end{equation}
	
	In the foregoing analysis we have assumed that the surface tension coefficient will vary linearly with the surface surfactant concentration, which is of the form
	\begin{equation}\label{surfacetension_linear}
		\sigma(\Gamma) = \sigma_0 - \sigma_\Gamma (\Gamma - \Gamma_E),
	\end{equation}
	where, $\Gamma_E$ is the equilibrium surface surfactant concentration and $\sigma_0 = \sigma(\Gamma_E)$ and $\sigma_\Gamma = - \left.\dfrac{\partial \sigma}{\partial \Gamma}\right|_{\Gamma = \Gamma_E}$. Equation (\ref{surfacetension_linear}) is only dependent on $\sigma_{\Gamma}$ and $\Gamma_E$, a generic specimen (prototype) equation of state, which is commonly used in thin film investigations to describe the solutal Marangoni effect \citep{kalliadasisThinFilmsSoft2007, oronLongscaleEvolutionThin1997}.  However, experiments have shown that surface tension can be approximated by a linear function throughout a wide concentration range \citep{gaverDropletSpreadingThin1992}.
	
	\subsection{\label{subsec:lubrication} Lubrication approximation and scaling}
	
	We chose a pair of separate characteristic length scales, $\hat{\lambda}$ in the longitudinal direction, whose order is equal to the disturbance wavelength and $\hat{h}_N$, the mean thickness of the film in the transverse direction. The velocity scale appears as $\hat{V}$. We will write more about the velocity scale later in the right place. In fact, we utilize the same scale as \cite{mukhopadhyayHydrodynamicInstabilityWave2021} to nondimensionalize the governing equations and boundary conditions. The extra set of nondimensional variables used in this work is $\Gamma^* = (1/\Gamma_\infty)\Gamma$ and $C^* = (\hat{h}/\Gamma_\infty)C$.

	The non-dimensional basic equations, after dropping the asterisk, for the problem are
	
	\begin{equation}\label{mass}
		\dfrac{\partial u}{\partial x} + \dfrac{\partial w}{\partial z} = 0,
	\end{equation}
	\begin{eqnarray}\label{x-momentum}
		\varepsilon Re\left(\dfrac{\partial u}{\partial t} + u \dfrac{\partial u}{\partial x} + w \dfrac{\partial u}{\partial z}\right) = - \varepsilon Re \dfrac{\partial p}{\partial x}+ \frac{Re}{Fr^2} + \left(\varepsilon^2 \dfrac{\partial^2 u}{\partial x^2} + \dfrac{\partial^2 u}{\partial z^2}\right),
	\end{eqnarray}
	\begin{eqnarray}\label{y-momentum}
		\varepsilon^2 Re\left(\dfrac{\partial w}{\partial t} + u \dfrac{\partial w}{\partial x} + w \dfrac{\partial w}{\partial z}\right) = -  Re \dfrac{\partial p}{\partial z} - \frac{Re}{Fr^2} \cot \theta  
		+ \left(\varepsilon^3 \dfrac{\partial^2 w}{\partial x^2} + \varepsilon \dfrac{\partial^2 w}{\partial z^2}\right),
	\end{eqnarray}
	\begin{equation}\label{concentration}
		\varepsilon Pe_b\left(\dfrac{\partial C}{\partial t} + u \dfrac{\partial C}{\partial x} + w \dfrac{\partial C}{\partial z}\right) = \varepsilon^2 \dfrac{\partial^2 C}{\partial x^2} + \dfrac{\partial^2 C}{\partial z^2}
	\end{equation}
	where, $Re = {\hat{V} \hat{h}_N}/{\nu}$, is the Reynolds number ($\nu = \eta/\rho$ is the kinematic viscosity), $Fr = \hat{V}/\sqrt{g \sin \theta \hat{h}_N}$ is the Froude number, $Pe_b = \hat{V}\hat{h}_N/D_b$, is the Péclet number, and $\varepsilon = 2\pi \hat{h}_N/\hat{\lambda} <<1$ is the aspect ratio.\\
	The boundary conditions at $z = 0$ are as follows:
	\begin{equation}\label{boundary_bottom}
		u = \beta \dfrac{\partial u}{\partial z}, \qquad w = 0, \qquad  \dfrac{\partial C}{\partial z} = 0,
	\end{equation}
	where $\beta = \sqrt{\mu}/(\upsilon \hat{h}_N)$ is the slip coefficient (dimensionless slip length) which measures the influence of the substrate permeability. Fluid with a base layer thickness of $3\times10^{-3}$m, an acceptable range of the dimensionless slip length range $(\beta)$ is $(0-0.08)$ makes sense \citep{samantaRoleSlipLinear2017}. \\
	The boundary conditions at $z = h$ are given by: 
	\begin{eqnarray}\label{boundary_normal}
		p_a - p + 2 Re^{- 1}\varepsilon \left\{\varepsilon^2 \dfrac{\partial u}{\partial x} \left(\dfrac{\partial h}{\partial x}\right)^2 - \left(\dfrac{\partial u}{\partial z} + \varepsilon^2  \dfrac{\partial w}{\partial x}\right) \dfrac{\partial h}{\partial x} + \dfrac{\partial w}{\partial z} \right\}
		\left(1 +\varepsilon^2 \left(\dfrac{\partial h}{\partial x}\right)^2\right)^{- 1}
		\nonumber\\
		= \varepsilon^2  \dfrac{\partial^2 h}{\partial x^2}\frac{We - Mn (\Gamma - \Gamma_e)}{\left\{1 +\varepsilon^2 \left(\dfrac{\partial h}{\partial x}\right)^2 \right\}^{3/2}}, \qquad
	\end{eqnarray}
	where $We = \sigma_0/(\rho \hat{V}^2 \hat{h}_N)$ is the Weber number,
	$Mn = \sigma_\Gamma \Gamma_\infty/(\rho \hat{V}^2 \hat{h}_N)$ is the Marangoni number,
	and $\Gamma_e = \Gamma_E/\Gamma_\infty$ is the dimensionless equilibrium surface surfactant concentration.
	Unless stated otherwise, parameter studies are performed by varying a single control parameter while holding all others fixed. In particular, when the equilibrium surface concentration $\Gamma_e$ is varied,
	the equation-of-state parameters $\sigma_0$ and $\sigma_\Gamma$ are kept fixed, so that the Weber number $We$ and Marangoni number $Mn$ remain independent
	control parameters. When $We$ or $Mn$ are instead recomputed from a $\Gamma_e$-dependent base state, this choice is stated explicitly.

	The rest of the boundary conditions at $z = h$ are:
	
	\begin{eqnarray}\label{boundary_tangential}
		\left(\dfrac{\partial u}{\partial z} + \varepsilon^2 \dfrac{\partial w}{\partial x}\right)\left\{1 - \varepsilon^2 \left(\dfrac{\partial h}{\partial x}\right)^2\right\}  - 4 \varepsilon^2 \dfrac{\partial u}{\partial x} \dfrac{\partial h}{\partial x}
		+ \varepsilon Re Mn \dfrac{\partial \Gamma}{\partial x} \left\{1 + \varepsilon^2 \left(\dfrac{\partial h}{\partial x}\right)^2\right\}^{1/2}= 0, \qquad
	\end{eqnarray}
	\begin{eqnarray}\label{boundary_kinematic}
		w = \dfrac{\partial h}{\partial t} +u \dfrac{\partial h}{\partial x},
	\end{eqnarray}
	
	\begin{eqnarray}\label{boundary_surface_sarfactant}
		\varepsilon \left[ \frac{\partial \Gamma}{\partial t} + \frac{1}{\sqrt{1 + \varepsilon ^2h_x^2}}  \frac{\partial (\Gamma u_s)}{\partial x} \right] =  \frac{\varepsilon ^2}{Pe_s\sqrt{1 + \varepsilon ^2h_x^2}} \frac{\partial}{\partial x} \left[ \frac{1}{\sqrt{1 + \varepsilon ^2 h_x^2}} \frac{\partial \Gamma}{\partial x} \right] + k_s[\kappa (1 - \Gamma) C - \Gamma]  \qquad
	\end{eqnarray}
	where, $Pe_s = \hat{V}\hat{h}_N/Ds$ surface Peclet number, $k_s = k_2 \hat{h}_N/\hat{V}$  is the desorption rate, $\kappa = k_1/(k_2\hat{h}_N)$ is the adsorption equilibrium constant,
	and $u_s(x,t) =\frac{1}{\sqrt{1 + \varepsilon^2h_x^2}} u(x, z = h,t)$ is the surface velocity. 
	
	To derive equation (\ref{boundary_surface_sarfactant}) from equation (\ref{surface_surfactant}) we have normalized $\nabla_s$ and $u_s$ by $\sqrt{1+ h_x^2}$,  the component of the fundamental tensor $g_{ij}$ on the free surface $z=h(x)$ in the streamwise direction\footnote{For alternative derivation we refer to Appendix A}.\par
	\citet{wongSurfactantMassBalance1996} developed a surfactant mass balance equation at a deforming fluid interface that considers both Eulerian and Lagrangian frames.

	We emphasize that the equation (\ref{boundary_surface_sarfactant}) is not the same as the corresponding equation in \citet{pascalStabilityInclinedFlow2019} and \citet{dalessioMarangoniInstabilitiesAssociated2020}. In effect, they have an additional term, namely $\varepsilon \Gamma \dfrac{\partial u}{\partial z} h_x$. 
	Within their reduced long-wave closure, this term is not accompanied by a corresponding contribution in the bulk transport equation, which, within that reduced closure, leads to a mismatch in the bulk-surface exchange terms and can compromise strict global surfactant mass conservation at the reduced-model level.
	As shown in Appendix~C, this manifests as a spurious long-time growth of the interfacial surfactant inventory.
	By contrast, Eq.~(\ref{boundary_surface_sarfactant}) reduces to the conservative surface-transport form of
	\citet{pereiraDynamicsFallingFilm2008} in the insoluble-surfactant limit.

	The remaining boundary condition in dimensionless form is as follows:
	\begin{eqnarray}\label{ficks}
		\dfrac{1}{Pe_b\left\{1 + \varepsilon^2 \left(\dfrac{\partial h}{\partial x}\right)^2\right\}^{1/2}} \left(\varepsilon^2 \dfrac{\partial h}{\partial x} \dfrac{\partial C}{\partial x}  - \dfrac{\partial C}{\partial z}\right) = k_s [\kappa (1 - \Gamma)C - \Gamma]
	\end{eqnarray}
	
	The Péclet number $Pe_{b,s}$ is a dimensionless ratio that compares the advection rate to the diffusion rate. The equilibrium constant of adsorption $\kappa$ is the ratio of the kinetic constants of the forward and backward reaction for the interaction between the interface and the bulk, providing a direct measure of the solubility of the surfactant in the bulk liquid \citep{jensenInsolubleSurfactantSpreading1992}. Surfactants with $\kappa \ll 1$ are highly soluble, while those with $ \kappa \gg 1 $ are sparingly soluble; that is, surfactants are trapped at the interface and nearly insoluble \citep{karapetsasPrimaryInstabilityFalling2013, kalogirouRoleSolubleSurfactants2019}. Furthermore, since $\kappa$ is the ratio between adsorption and desorption rate, $\kappa \to \infty$  corresponds to an insoluble surfactant (as for the insoluble limit $k_{1,2} \to 0$).

	\subsection{\label{subsec:base} Base profiles}

	The steady flow of the equilibrium solution corresponding to the uniform stream direction (base state) can be found by setting $\dfrac{\partial}{\partial t} \equiv 0$, $\dfrac{\partial}{\partial x} \equiv 0$ and $w = 0$.
	\begin{equation}
		U_B(z) = \frac{Re}{Fr^2} \left(-\frac{1}{2} z^2 + z + \beta\right), \qquad W_B = 0,
	\end{equation}
	\begin{equation}
		P_B(z) = - \frac{Re}{Fr^2} \left(\frac{\cot\theta}{Re}\right)(z -1), \qquad h_B= h_e = 1
	\end{equation}
	\begin{equation}
		C_B \equiv C_e = \frac{\Gamma_e}{\kappa(1 - \Gamma_e)}, \qquad \Gamma_B = \Gamma_e,
	\end{equation}
	where, $C_e$ and $\Gamma_e$ are the  bulk and surface surfactant concentration respectively at the equilibrium state, which is also called the uniform base level concentration. Since the surface concentration has been scaled by its maximal value, the range of the control parameter $\Gamma_e$ will be $0\le \Gamma_e \le 1$. To achieve a finite bulk concentration, $\kappa \to \infty$ as $\Gamma_e \to 1^-$, ensuring that the product $\kappa(1 - \Gamma_e)$ does not approach zero. Furthermore, at this moment, the surface surfactant concentration $\Gamma_e$ and the bulk surfactant concentration $C_e$ are in equilibrium with $J_{bs}$ = 0, and the equation (\ref{Langmuir}) produces the following relationship: $C_e = \Gamma_e/\kappa(1-\Gamma_e)$.  It should be noted that when $C_e > 1$, the bulk surfactant forms micelles.  This study focuses exclusively on the properties of soluble surfactants, including solubility, sorption kinetics, and bulk diffusivity, when $C_e < 1$ \citep{liRoleSolubleSurfactant2023}. For insoluble surfactant $\kappa \to \infty$, we anticipate that the concentration of surfactant in the bulk will vanish at the insoluble limit. This implies that maximum surface concentration can only be achieved if the surfactant is insoluble. 
	
	\begin{figure} 
		\centering
		\subcaptionbox{The variation of the base flow}
		{\includegraphics[width=0.49\textwidth]{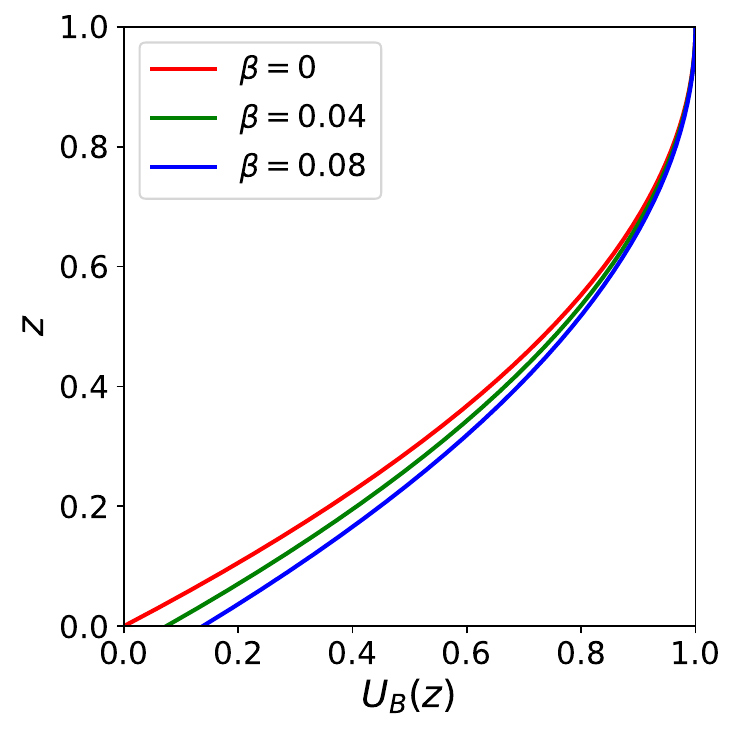}}%
		\hfill 
		\subcaptionbox{The variation of the wall shear stress}
		{\includegraphics[width=0.49\textwidth]{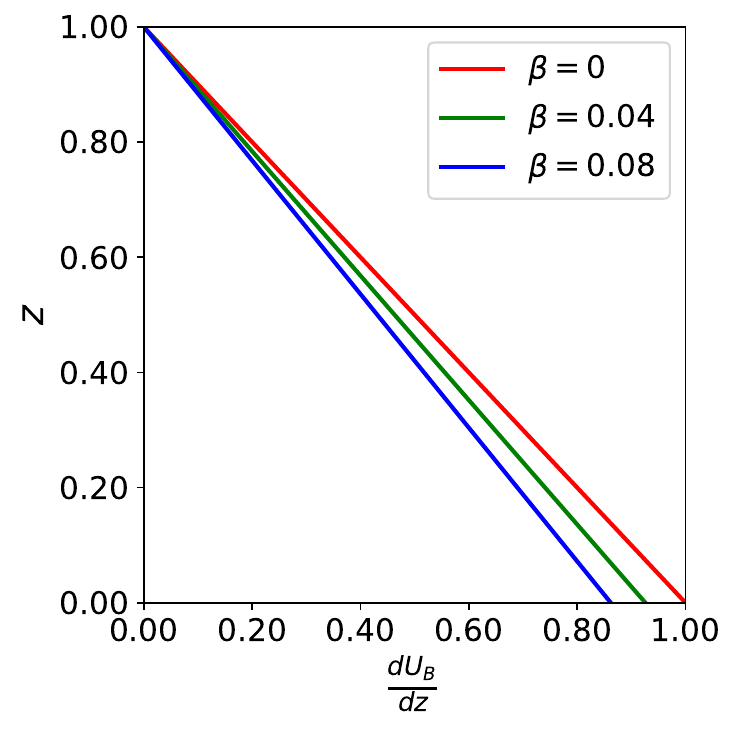}}
		\hfill 
		\caption{Base flow and wall shear stress for different values of the slip parameter $\beta$, the velocity scale is twice the free surface velocity.}
		\label{fig:base_flow}
	\end{figure}

	The base flow solution shows that the flow velocity $U_B (z)$ is a quadratic function of $z$ with a semiparabolic form, peaks at the film surface.  It should be emphasized that the base velocity is independent of the surfactant concentration.  Figure \ref{fig:base_flow}(a) shows that the base velocity in the stream direction increases in the presence of the slip length \citep{mukhopadhyayHydrodynamicInstabilityWave2021}.  In contrast, the slip length parameter reduces the wall shear stress, as seen in Figure \ref{fig:base_flow}(b).  This result confirms that the slippery plane has a significant impact on the base flow velocity.

	\section{Orr-Sommerfeld (OS) Boundary Value Problem}
	We linearize the complete system around the base state and introduce perturbations in normal mode. Since we know a normal mode analysis requires splitting each unknown $X$ into $X = X_B + \tilde{X} \exp[ik(x - \alpha t)]$. Where $X_B$ represents the base flow, while $\tilde{X}(z)$ represents the infinitesimal amplitude of a sine wave with wave number $k$ and celerity $\alpha$. Therefore, we linearize the governing equations (\ref{mass}-\ref{ficks}) and applying normal modes yields the Orr-Sommerfeld eigenvalue problem \citep{yihStabilityLiquidFlow1963}.
	\begin{eqnarray} \label{orr}
		(D^2-k^2)^2 \psi - {i}\: k Re \left[\left(U_B-\alpha\right)(D^2-k^2)\psi-{U_B}'' \psi\right] = 0,\quad
	\end{eqnarray}	
	
	\begin{eqnarray} \label{surf}
		(D^2 - k^2) \mathtt{c} - i k Pe_b (U_B - \alpha) \mathtt{c} = 0,
	\end{eqnarray}		
	where $\psi(z)$ and $\mathtt{c}(z)$ are the amplitude of the stream function and the surfactant concentration, respectively, and $D \equiv d/dz$. The associated boundary conditions are as follows:
	\begin{equation}\label{bcorrnp}
		\psi(0) = 0,
	\end{equation}
	\begin{equation}\label{bcorrs}
		\beta \psi''(0) - \psi'(0) = 0,
	\end{equation}	
	\begin{equation}\label{bcorrcon}
		\mathtt{c}'(0) = 0,
	\end{equation}
	\begin{equation}\label{bcorrt}
		\psi''(1) + k^2\psi(1) + i \:k Re Mn \gamma -  \frac{{U_B}''(1) \psi(1)}{[  {U_B}(1) - \alpha]} = 0,
	\end{equation}
	
	\begin{eqnarray}\label{bcorrn}
		\psi'''(1) - 3 k^2 \psi'(1) - i\: Re \: k[\psi'(1)(U_B(1)-\alpha)] 
		\nonumber\\
		+ i \: k \left(\frac{\left(\frac{Re}{Fr^2}\right)\cot \theta + k^2 Re We}{ U_B(1) - \alpha}\right)\psi(1) = 0,
	\end{eqnarray}
	
	\begin{eqnarray}
		\gamma k^2 + i k Pe_s [\Gamma_B\psi'(1) - \gamma(\alpha - {U_B}(1))]
		\nonumber\\
		+ k_s Pe_s[\gamma (1 + \kappa C_B(1)) - \kappa 	\mathtt{c}(1)(1 - \Gamma_B) ] = 0,
	\end{eqnarray}
	\begin{eqnarray}
		\mathtt{c}'(1) =  Pe_b k_s \left[\gamma (1 + \kappa C_B(1)) - \kappa \mathtt{c}(1)(1 - \Gamma_B) \right],
	\end{eqnarray}
	where, $\gamma$ denotes the amplitude of the perturbed part of the surface surfactant concentration, and the primes indicate differentiation with respect to $z$. Since the Orr-Sommerfeld framework applies to disturbances of arbitrary wavelength, the long-wave parameter $\varepsilon$ is removed by introducing the scaled variables 
	$\hat{x} = x/ \varepsilon$ and  $\hat{t} = t/\varepsilon$, and subsequently omitting the hat ($\hat{.}$) notation. This rescaling ensures identical characteristic length scales in the streamwise and normal directions, thereby extending the analysis beyond the long-wave approximation.
	Note that the Orr-Sommerfeld problem is written in the rescaled variables used for arbitrary wavelengths, so the long-wave aspect ratio $\varepsilon$ is absorbed into the streamwise and time scalings.
	Parameters defined later within the reduced long-wave model (e.g., $Ka=\varepsilon^2 Re\,We$) therefore do not
	appear explicitly in Eqs.~(\ref{orr})-(\ref{surf}), but enter in the long-wave/DA model.

	\subsection{Asymptotic solution of the OS problem} 
	To predict the onset of instability, we apply an asymptotic analysis as $k \rightarrow 0$. This is because perturbations of small wavenumbers are more unstable. Long-wavelength asymptotic analysis can solve the eigenvalue problem by expanding the eigenfunction/flow variables $\Psi$, $c$, $\gamma$ and the eigenvalues $\alpha$ in powers of $k$ \citep{yihStabilityLiquidFlow1963}.\\
	
	$\psi(z) = \psi_0(z) + k \psi_1(z) + O(k^2),\quad
	\mathtt{c}(z) = \mathtt{c}_0(z) + k \mathtt{c}_1(z) + O(k^2), \quad \gamma(z) = \gamma_0(z) + k \gamma_1(z) + O(k^2)$ \quad  and $ \alpha = \alpha_0 + k \alpha_1 + O(k^2)$. \par
	
	Solving the hierarchy of problems at various orders of $k$, we can find the complex phase speed $\alpha = {\alpha}_r + i {\alpha}_i =  {\alpha}_0 + k  {\alpha}_1$\footnote{This eigen value corresponds to the hydrodynamic Kapitza mode $\alpha^{(1)}$; for more information, see Appendix B.}, where, 
	
	\begin{eqnarray} \label{phase_speed}
		\alpha_0 = (1 + 2 \beta)\dfrac{Re}{Fr^2}
	\end{eqnarray}
	
	\begin{eqnarray}
		\alpha_1 = i Re \left[\dfrac{(1 + \beta)\lbrace2 + 5\beta(2 + 3\beta)\rbrace}{15}\left(\dfrac{Re}{Fr^2}\right)^2 - \dfrac{(1 + 3 \beta)}{3}\left(\dfrac{Re}{Fr^2}\right) \dfrac{\cot\theta}{Re}\right.
		\nonumber\\
		\left. - \dfrac{3 (1 + 2 \beta)Mn \Gamma_e \kappa(1-\Gamma_e)^2}{4 + 3 \kappa(1 - \Gamma_e)^2} \right]
	\end{eqnarray}
	
	The neutral stability is given by $\alpha_1 = 0$, which gives
	
	\begin{eqnarray}
		\dfrac{\cot\theta}{Re} = \dfrac{2 + 5 \beta(2 + 3\beta)}{5} \left(\dfrac{Re}{Fr^2}\right) 
		\nonumber\\
		\left[\left(\dfrac{1 + \beta}{1 + 3\beta}\right) - \dfrac{45 (1 + 2\beta)}{(1 + 3\beta)\lbrace 2 + 5\beta (2 + 3\beta)\rbrace} \left(\dfrac{Re}{Fr^2}\right)^{-2} \dfrac{Mn \Gamma_e \kappa (1 - \Gamma_e)^2}{4 + 3\kappa(1 - \Gamma_e)^2}\right]
	\end{eqnarray}
	
	The onset of instability is obtained by considering the zero critical wave number \citep{scheidThermocapillaryLongWaves2005}. This yields the critical condition
	
	\begin{eqnarray} \label{critical_soluble}
		Re_c = \underbrace{ \dfrac{5}{\lbrace 2 + 5 \beta (2 + 3\beta)\rbrace} \left(\dfrac{1 + 3 \beta}{1 + \beta}\right) \left(\dfrac{Re}{Fr^2}\right)^{-1} \cot\theta}_\text{Gravity effect}
		\nonumber\\
		+ \underbrace{ \dfrac{45 (1 + 2\beta)}{(1 + \beta)\lbrace 2 + 5\beta (2 + 3\beta)\rbrace} \left(\dfrac{Re}{Fr^2}\right)^{-2}\left[ \dfrac{Mn \Gamma_e \kappa (1 - \Gamma_e)^2}{4 + 3\kappa(1 - \Gamma_e)^2}\right]}_\text{Surfactant effect}
	\end{eqnarray} 
	In the insoluble limit, letting $\kappa \to \infty$, we have
	\begin{eqnarray} \label{critical_insoluble}
		Re_c =  \dfrac{5}{\lbrace 2 + 5 \beta (2 + 3\beta)\rbrace} \left(\dfrac{1 + 3 \beta}{1 + \beta}\right) \left(\dfrac{Re}{Fr^2}\right)^{-1} \cot\theta 
		\nonumber\\
		+   \dfrac{15 (1 + 2\beta)}{(1 + \beta)\lbrace 2 + 5\beta (2 + 3\beta)\rbrace} \left(\dfrac{Re}{Fr^2}\right)^{-2} Mn \Gamma_e  
	\end{eqnarray} 
	To make a comparison to the previously established work, we must use proper scaling, for example: 
	\begin{enumerate}
		\item The choice is of the velocity scale $\hat{V} = \langle u \rangle = g \sin \theta \hat{h}_N^2/3\nu$ the average Nusselt velocity for flat rigid plate, we get $Re/Fr^2 = (g\sin\theta \hat{h}_N^2/\nu)(1/\hat{V}) = 3$ \citep{pascalLinearStabilityFluid1999}. 
		\item For,  $\hat{V} = 2 \hat{U}_N = (g \sin \theta \hat{h}^2_N/\nu)(1 + 2 \beta)$ as twice the dimensional Nusselt free surface velocity and redefine the Reynolds number as $Re = \hat{U}_N \hat{h}_N /\nu$, then $Re/Fr^2 = 2/(1 + 2\beta)$\citep{samantaFallingFilmSlippery2011}.
		\item If we choose the viscous gravity length scale $l_\nu = \nu^{2/3} (g \sin \theta)^{-1/3}$ and $t_\nu = \nu^{1/3} (g \sin \theta)^{-2/3}$ and the velocity scale $\hat{V} = U_N = (g\sin\theta \hat{h}_N^2/\nu)(1 + 2 \beta)$, we get $Re/Fr^2 = 1 /(1 + 2 \beta)$ \citep{kalliadasisFallingLiquidFilms2012}. 
	\end{enumerate}
	
	For detailed information, see \citet{mukhopadhyayHydrodynamicInstabilityWave2021}. 
	Therefore, if we choose the velocity scale as the twice the dimensional Nusselt free surface velocity and redefine Reynolds number as $Re = \hat{U}_N \hat{h}_N /\nu$ then $Re/Fr^2 = 2/(1 + 2\beta)$, Equations (\ref{critical_soluble}) and (\ref{critical_insoluble}) reduce to: 
	\begin{equation} \label{critical_soluble_scal}
		Re_c =   \dfrac{5(1+ 2 \beta)}{2\lbrace 2 + 5 \beta (2 + 3\beta)\rbrace} \left(\dfrac{1 + 3 \beta}{1 + \beta}\right) \cot\theta +  \dfrac{45 (1 + 2\beta)^3}{4(1 + \beta)\lbrace 2 + 5\beta (2 + 3\beta)\rbrace} \left[ \dfrac{Mn \Gamma_e \kappa (1 - \Gamma_e)^2}{4 + 3\kappa(1 - \Gamma_e)^2}\right] 
	\end{equation}   
	and
	\begin{equation}\label{critical_insoluble_scal}
		Re_c =  \dfrac{5(1+2\beta)}{2 \lbrace 2 + 5 \beta  (2 + 3\beta)\rbrace} \left(\dfrac{1 + 3 \beta}{1 + \beta}\right) \cot\theta  +   \dfrac{15 (1 + 2\beta)^3}{4(1 + \beta)\lbrace 2 + 5\beta (2 + 3\beta)\rbrace} Mn \Gamma_e  
	\end{equation}  
	There are numerous limits and comparisons that can be made. To obtain the no-slip limit, set $\beta = 0$. Equations (\ref{critical_soluble_scal}) and (\ref{critical_insoluble_scal}) reduces to 
	\begin{equation}\label{critical_soluble_noslip}
		Re_c =  \dfrac{5}{4} \cot\theta + \dfrac{45} {8} \left[ \dfrac{Mn \Gamma_e \kappa (1 - \Gamma_e)^2}{4 + 3\kappa(1 - \Gamma_e)^2}\right]    
	\end{equation}  
	and
	\begin{equation}\label{critical_insoluble_noslip}
		Re_c =  \dfrac{5}{4} \cot\theta +\dfrac{15} {8} Mn \Gamma_e,
	\end{equation} 
	respectively. The expression in Equation (\ref{critical_soluble_noslip}) is identical to that reported by \citet{karapetsasRoleSurfactantsMechanism2014} in the context of the isothermal limit of their study. Equations (\ref{critical_soluble_noslip}) and (\ref{critical_insoluble_noslip}) are identical to Equations (97) and (95) (in the context of isothermal flow), respectively, as reported by \citep{samantaEffectSolubleSurfactant2025}. \\
	
	For the non-surfactant limit with the average Nusselt velocity for flat rigid plate as the velocity scale, we have
	
	\begin{eqnarray}
		Re_c = \dfrac{5}{3 \lbrace 2 + 5 \beta (2 + 3\beta)\rbrace} \left(\dfrac{1 + 3 \beta}{1 + \beta}\right)\cot\theta.
	\end{eqnarray}
	This is exactly the same as \citet{sadiqThinNewtonianFilm2008}.
	
	To isolate the effect of wall slip on the onset of instability, we first examine how the critical Reynolds number $Re_c$ varies with the slip parameter $\beta$ for different equilibrium surface coverages. This representation cleanly separates changes in the linear
	stability threshold from changes in nonlinear wave morphology that arise solely from differing distances to onset. 
	\begin{figure}
		\centering
		\includegraphics[width=1\linewidth]{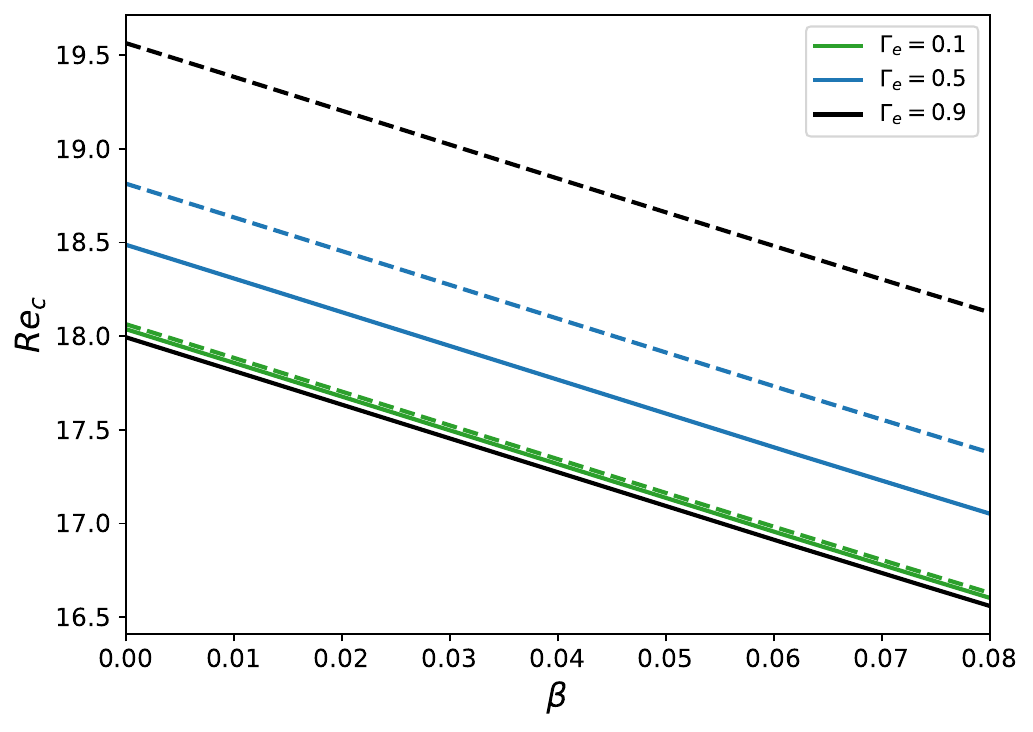}
		\caption{Critical Reynolds number $Re_c$ as a function of the slip parameter $\beta$ for different base-state surfactant concentrations $\Gamma_e$. Solid lines correspond to a soluble surfactant, while dashed lines show the corresponding insoluble-surfactant cases. Colors indicate the value of $\Gamma_e$.}
		\label{fig:Rec_beta}
	\end{figure} 
	Figure~\ref{fig:Rec_beta} shows that the onset threshold depends sensitively on both the slip parameter $\beta$
	and the base-state surfactant concentration. For a fixed $\Gamma_e$, increasing $\beta$ modifies the wall shear and the base velocity profile, which changes the balance between gravitational forcing and surfactant-induced Marangoni stresses. The comparison between solid and dashed curves indicates that surfactant solubility alters this balance by allowing additional redistribution of surfactant through bulk–interface exchange. As a result, the effect of slip on $Re_c$ is not identical for soluble and insoluble surfactants, even when the
	interfacial concentration is the same. The separation between solid and dashed curves at fixed $\Gamma_e$ directly shows that wall slip and surfactant
	solubility do not act independently on the onset threshold.
	\par In the nonlinear analysis below, the critical values $Re_c$ obtained here are used to define the distance from onset,
	allowing parameter effects to be compared at fixed supercriticality.

	\begin{figure} 
		\centering
		\subcaptionbox{$\beta = 0.04, Mn = 1, \kappa = 10, \theta = 4^\circ$}
		{\includegraphics[width=0.49\textwidth]{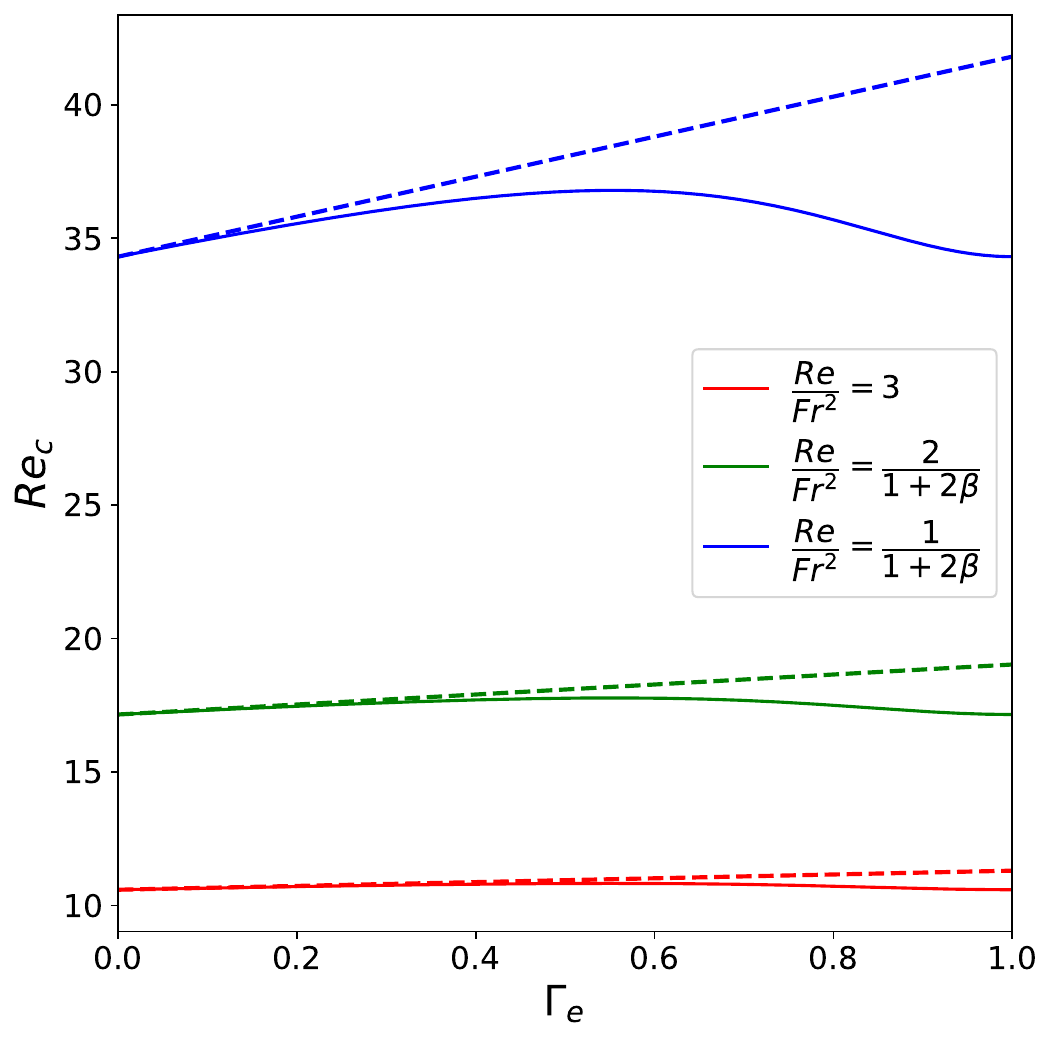}}%
		\hfill 
		\subcaptionbox{$\beta = 0.04, Mn = 1, \dfrac{Re}{Fr^2} = \dfrac{2}{1 + 2 \beta}, \theta = 4^\circ$}
		{\includegraphics[width=0.49\textwidth]{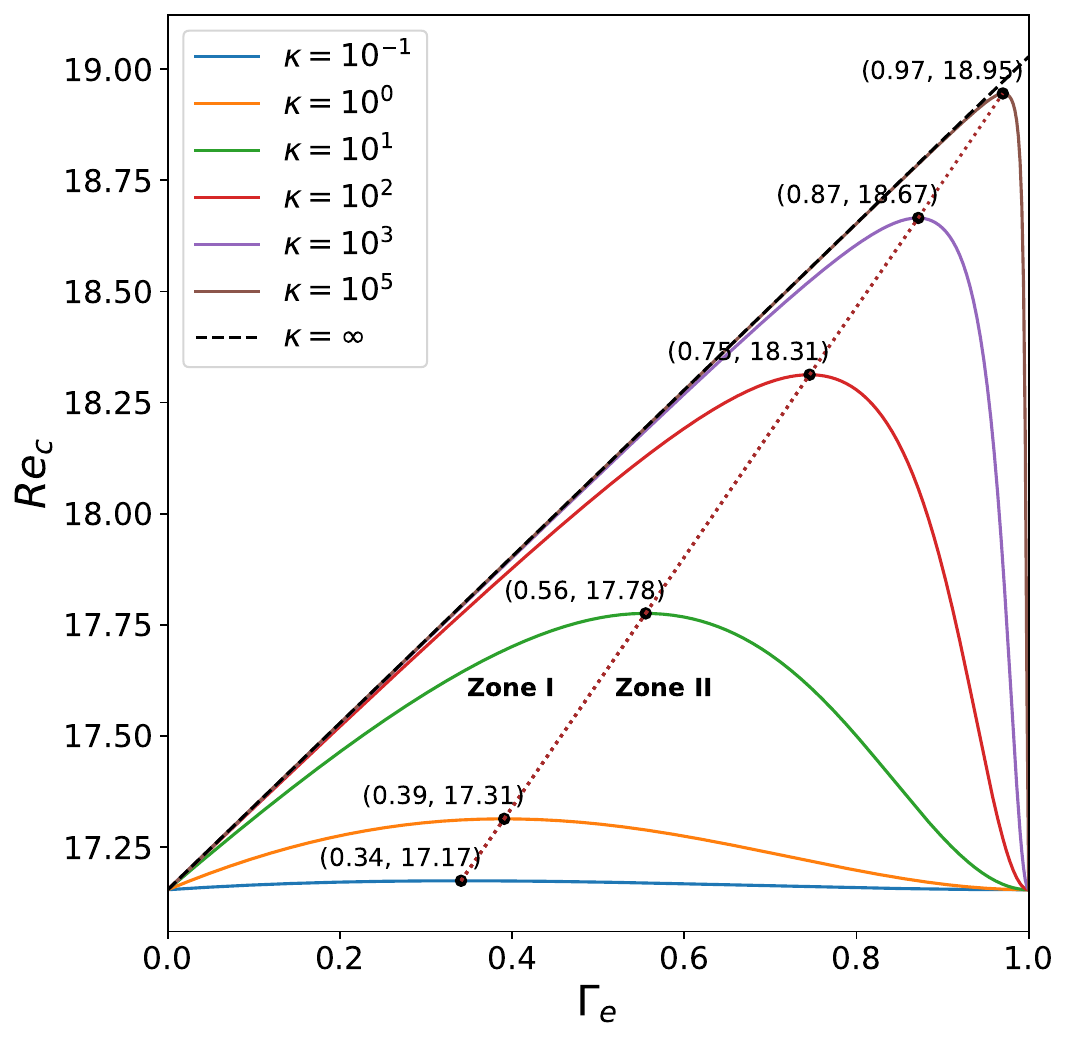}}
		\hfill 
		\subcaptionbox{$\kappa = 10, Mn = 1, \dfrac{Re}{Fr^2} = \dfrac{2}{1 + 2 \beta}, \theta = 4^\circ$}
		{\includegraphics[width=0.49\textwidth]{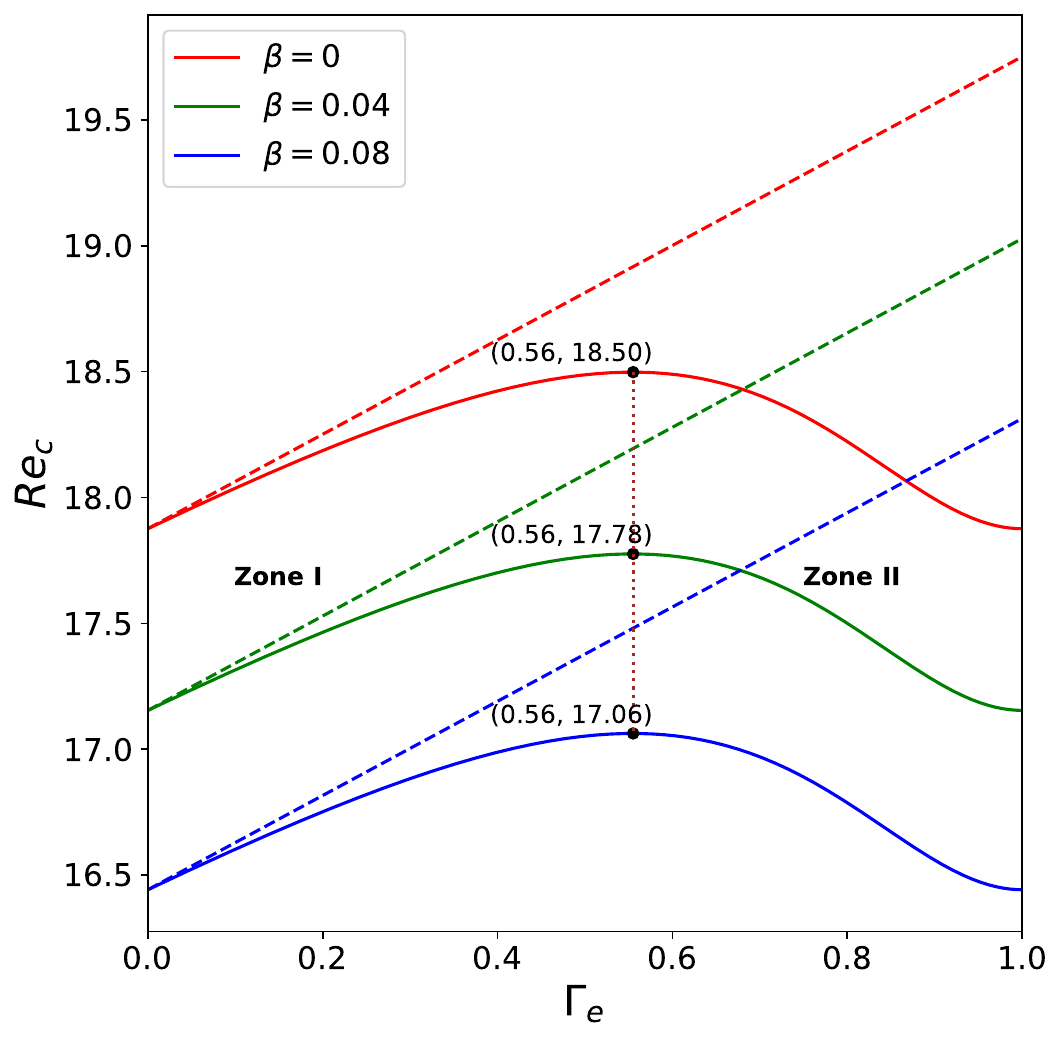}}
		\hfill 
		\subcaptionbox{$\beta = 0.04, \kappa = 10, \dfrac{Re}{Fr^2} = \dfrac{2}{1 + 2 \beta}, \theta = 4^\circ$}
		{\includegraphics[width=0.49\textwidth]{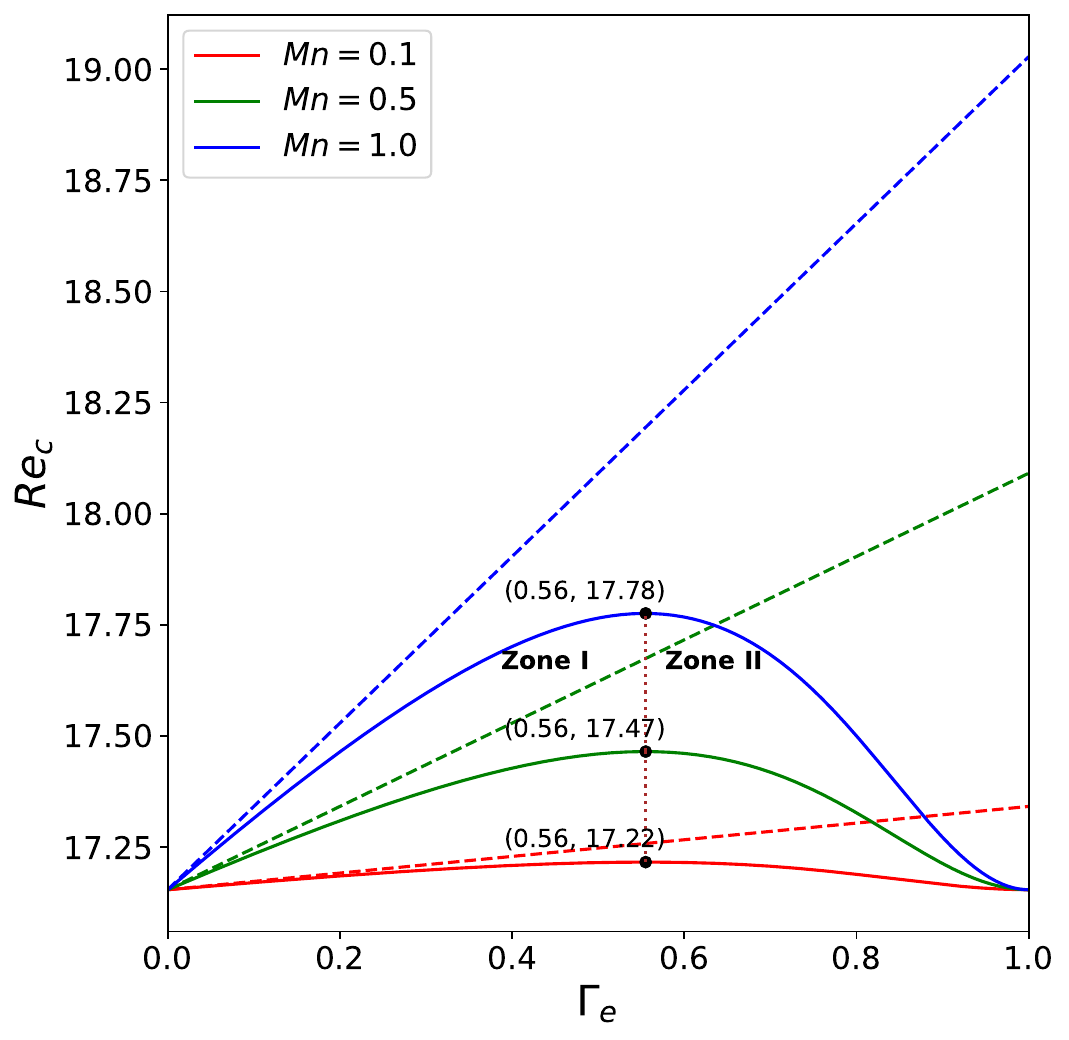}}
		\hfill 
		\caption{Effect of different parameters in critical Reynolds as a function of $\Gamma_e$. The solid lines represent the results for soluble surfactant, whereas the dashed line represents respective cases for insoluble surfactant. The insoluble cases are plotted using the equation (\ref{critical_insoluble}).}
		\label{fig:linear_analytic}
	\end{figure}

	Figure (\ref{fig:linear_analytic}) shows the variation of the critical Reynolds number $Re_c$ with the equilibrium surface surfactant parameter $\Gamma_e$. The results show that for insoluble surfactants ($\kappa \to \infty$ dashed line in Figure (\ref{fig:linear_analytic}) (b)), $Re_c$ increases steadily with $\Gamma_e$, indicating a stabilizing influence of $\Gamma_e$ under all circumstances. However, the picture differs for soluble surfactants. In the case of soluble surfactants, $Re_c$ grows with $\Gamma_e$, reaches its maximum, and then declines, resulting in two zones, reflecting its double role. It provides stability in the first zone before the maximum and instability in the second zone thereafter. The dual role of $\Gamma_e$ can be articulated in accordance with the elucidation provided by \citet{georgantakiMeasurementsStabilisationLiquid2016}. Increasing $\Gamma_e$ leads to an increase in the gradient of surface surfactant concentration, or surface elasticity.  As a result, the generated Marangoni stress reduces the surface deformation and stabilizes the instability of the mode.  The local equilibrium near the free surface of the film causes the concentration of the bulk surfactant to increase, resulting in a higher mass exchange between the surface and the concentration of the bulk surfactant and an intensified concentration gradient across the surface.  As a result, the influence of surface elasticity is eliminated, and the film surface behaves similarly to a smooth viscous surface. It is important to note that for different values of $\kappa$ (panel (b)), we find that the maximum $Re_c$ shifts to $\Gamma_e \approx 1$ in a non-linear fashion as the parameter $\kappa$ increases showing its stabilizing impact, but for $Mn$, the line of maximum $Re_c$ is almost vertical (panel (d)) and its stabilizing impact is not as strong as $\kappa$. The scenario of an insoluble surfactant-impure film flow is more stable than that of the soluble surfactant-laden film flow because, physically, increasing $\kappa$ indicates a stronger adsorption of surfactant at the film surface than desorption of surfactant.  Lastly, we observe that the present findings are in agreement with those of \citet{karapetsasRoleSurfactantsMechanism2014} and \citet{samantaEffectSolubleSurfactant2025}.
	The slip parameter $\beta$ gives a destabilizing effect (panel (c)). Panel (a) shows the effects of scaling, which depend only on the observer.

	\subsection{Numerical solution of OS problem}
	In order to determine the linear stability of perturbations of an arbitrary wavelength, we obtain numerical
	solutions to the Orr-Sommerfeld eigenvalue problem. \citet{gottliebNumericalAnalysisSpectral1977} proposed the Chebyshev method along with the QZ algorithm. In this method, each flow variable is expanded as a series of Chebyshev polynomials.
	\begin{equation}\label{chebpoly}
		\psi(\xi) =  \sum_{j=0}^{N} a_j T_j(\xi), \quad \quad
		\mathtt{c}(\xi) = \sum_{j=0}^{N} b_j T_j(\xi) 
	\end{equation}
	where, $a_j$'s, and  $b_j$'s are the discrete Chebyshev expansion coefficients \citep{trefethenSpectralMethodsMATLAB2000} to be  determined from the numerical simulation and the transformation $\xi = 2 z - 1$ is used to shift the liquid layer domain [0, 1] to the computational domain [-1, 1] in which the Chebyshev polynomials $T_j(\xi)$ are defined, and 
	
	$$T_j(\xi) = \cos(j\cos^{-1}(\xi))$$
	Substituting (\ref{chebpoly}) in the governing equations, one can recast them into a generalized matrix eigenvalue problem
	\begin{equation} \label{matrix}
		\mathcal{A} \wp = \alpha \mathcal{B} \wp,   
	\end{equation}
	
	where $\wp = [\psi_0 \quad \psi_1 \quad ... \quad \psi_N \quad \mathtt{c}_0 \quad \mathtt{c}_1 \quad ... \quad \mathtt{c}_N\quad \gamma]$ is the eigenvector ($\psi_i=\psi(\xi_i)$, $\mathtt{c}_i=\mathtt{c}(\xi_i)$),
	$\mathcal{A}$ and $\mathcal{B}$ are $(2N + 3) \times (2N + 3)$ matrices, 
	and eigenvalues $\alpha$ provide all possible temporal modes. The most unstable temporal mode predicts the instability of the given flow configuration. In the present work, we have used the Chebyshev spectral collocation method proposed by \citet{henningsonStabilityTransitionShear2012}. In this method, the Chebyshev polynomials are evaluated at the Gauss-Lobatto collocation points $\xi_i = - \cos (i \pi /N) \quad (i = 0, .., N)$, which are actually the extrema of the Chebyshev polynomials.
	
	\begin{figure} 
		\centering
		\subcaptionbox{}
		{\includegraphics[width=0.49\textwidth]{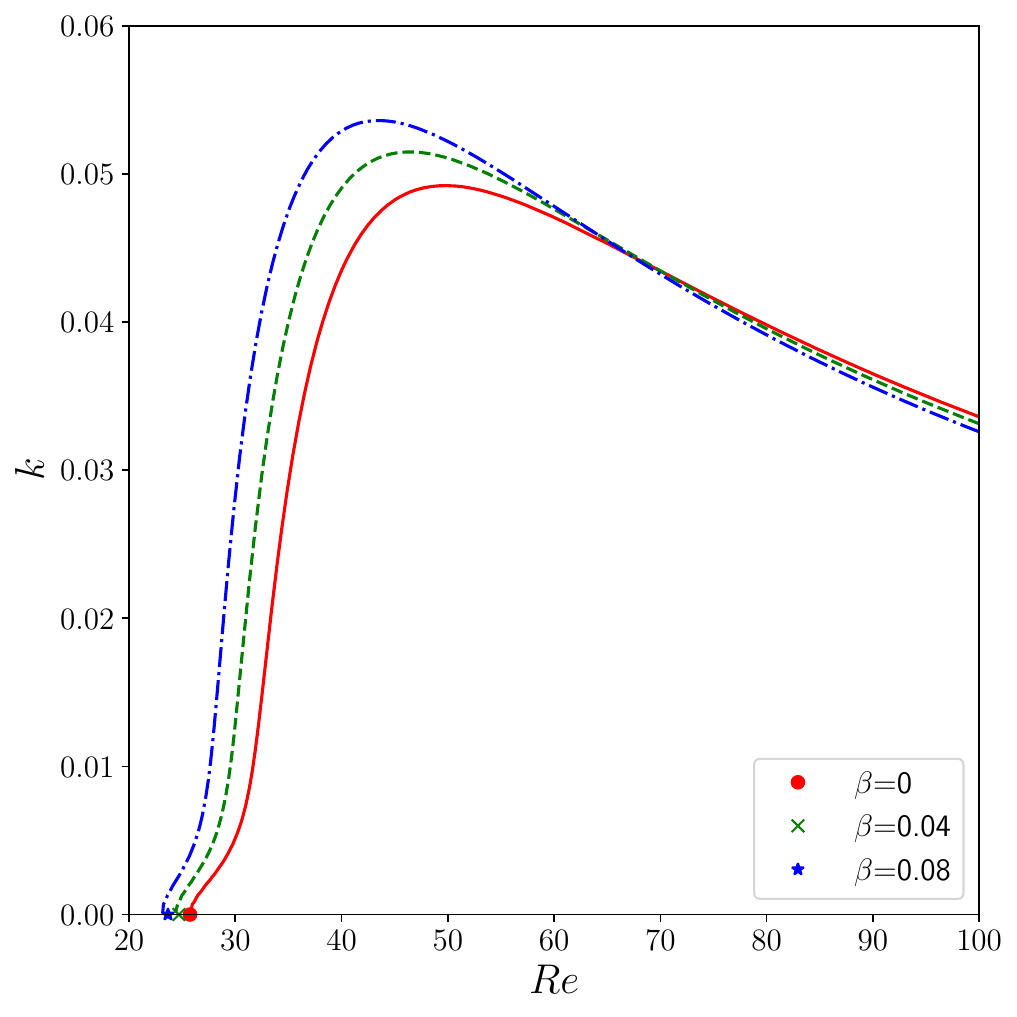}}%
		\hfill 
		\subcaptionbox{$Re = 40$}
		{\includegraphics[width=0.49\textwidth]{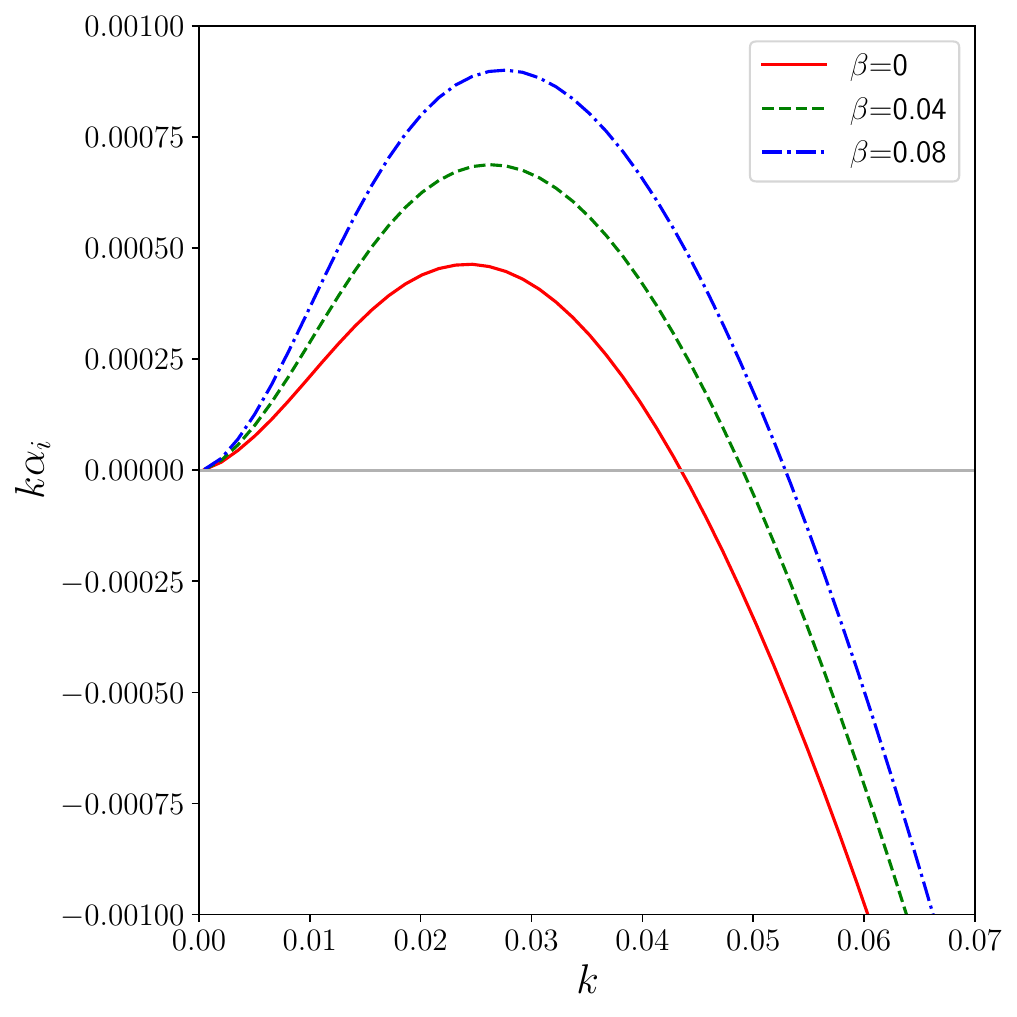}}
		\hfill 
		\subcaptionbox{$Re = 100$}
		{\includegraphics[width=0.49\textwidth]{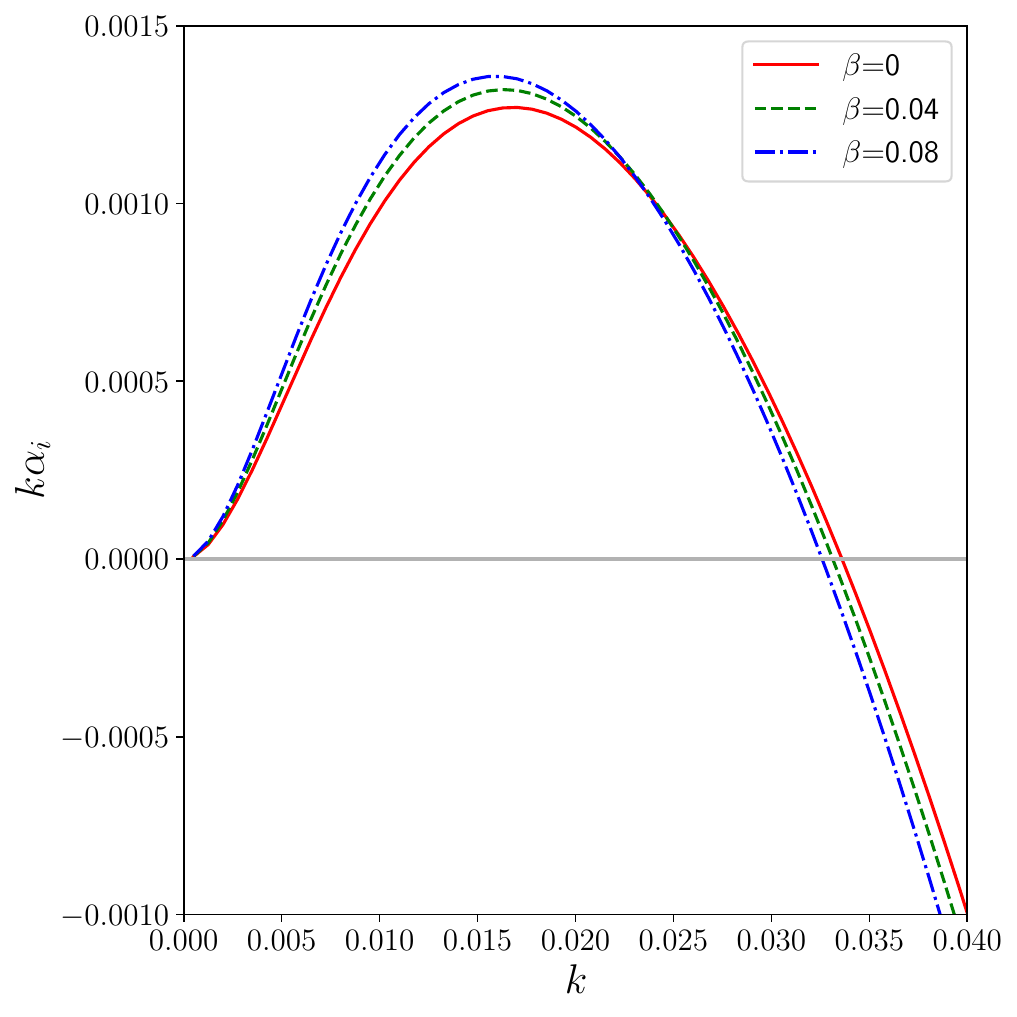}}
		\hfill 
		\subcaptionbox{$Re = 200$}
		{\includegraphics[width=0.49\textwidth]{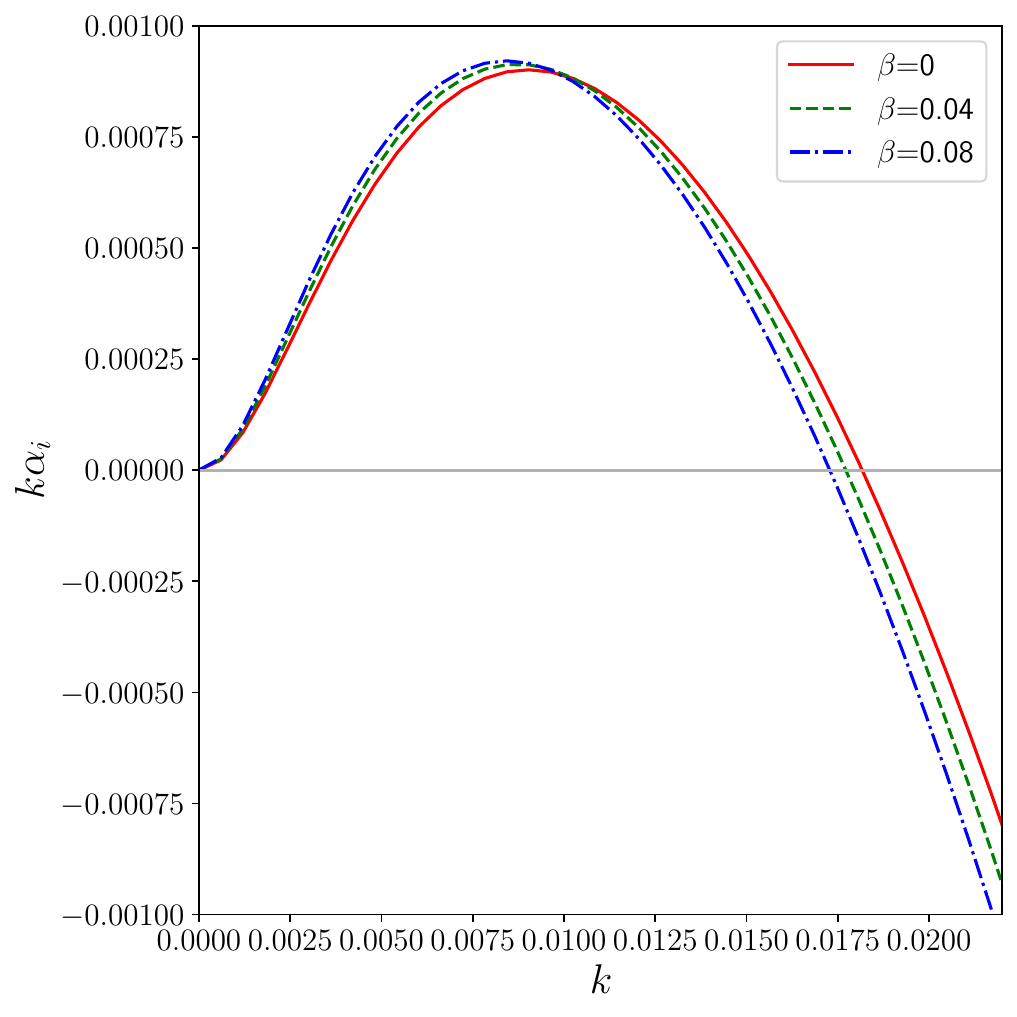}}
		\hfill 
		\caption{Effect of slip parameter ($\beta$) in linear stability. $Mn=0.5, \Gamma_e=0.5,\kappa = 10, Pe_b = Pe_s = 700, k_s = 0.5, \theta = 4^{\circ}, Ka=1000$. Left panel: marginal stability curve, right upper panel: growth rate curve for $Re = 40$, right lower panel: growth rate curve for $Re = 100$.  The marker in panel (a) represents the critical Reynolds number obtained through asymptotic analysis, but the other legends remain consistent.}
		\label{fig:liear_influence_beta}
	\end{figure}

	\begin{figure} 
		\subcaptionbox{}
		{\includegraphics[width=0.49\textwidth]{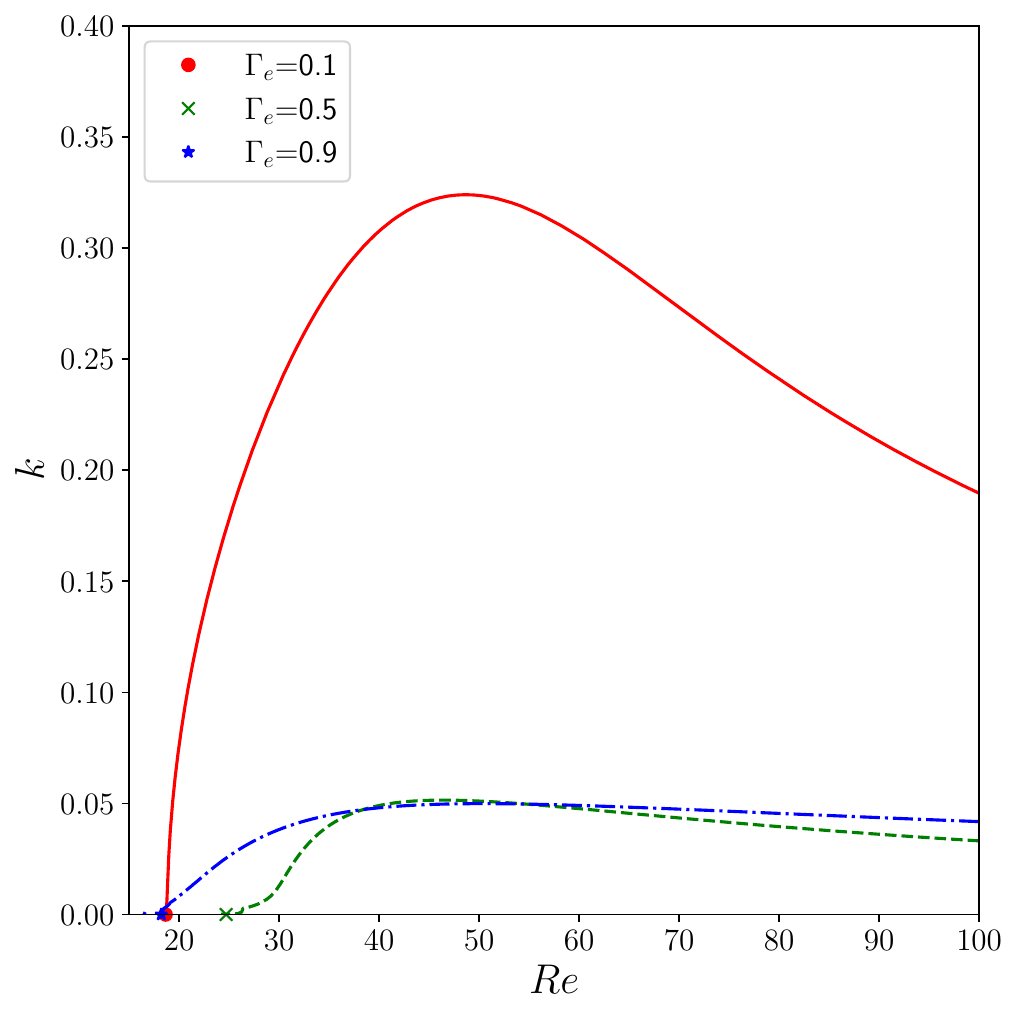}}%
		\hfill 
		\subcaptionbox{$Re = 40$}
		{\includegraphics[width=0.49\textwidth]{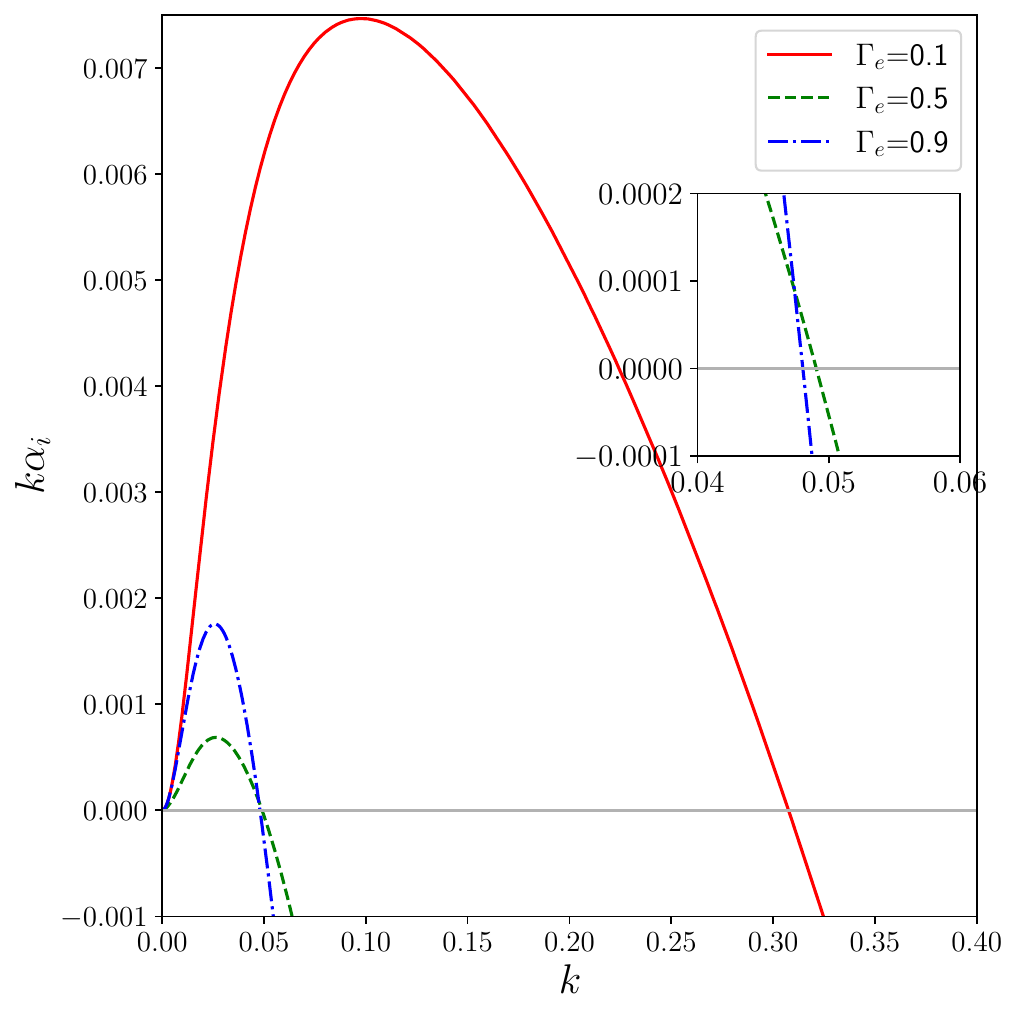}}
		\hfill 
		\caption{Effect of $\Gamma_e$ in linear stability. $Mn=0.5, \beta=0.04,\kappa = 10, Pe_b = Pe_s=700, k_s=0.5, \theta = 4^{\circ}, Ka=1000$. Left panel: marginal stability curve, right panel: growth rate curve for $Re = 40$. The marker in panel (a) represents the critical Reynolds number obtained through asymptotic analysis, but the other legends remain consistent.}
		\label{fig:liear_influence_Gamma_e}
	\end{figure}
	
	\begin{figure} 
		\subcaptionbox{}
		{\includegraphics[width=0.49\textwidth]{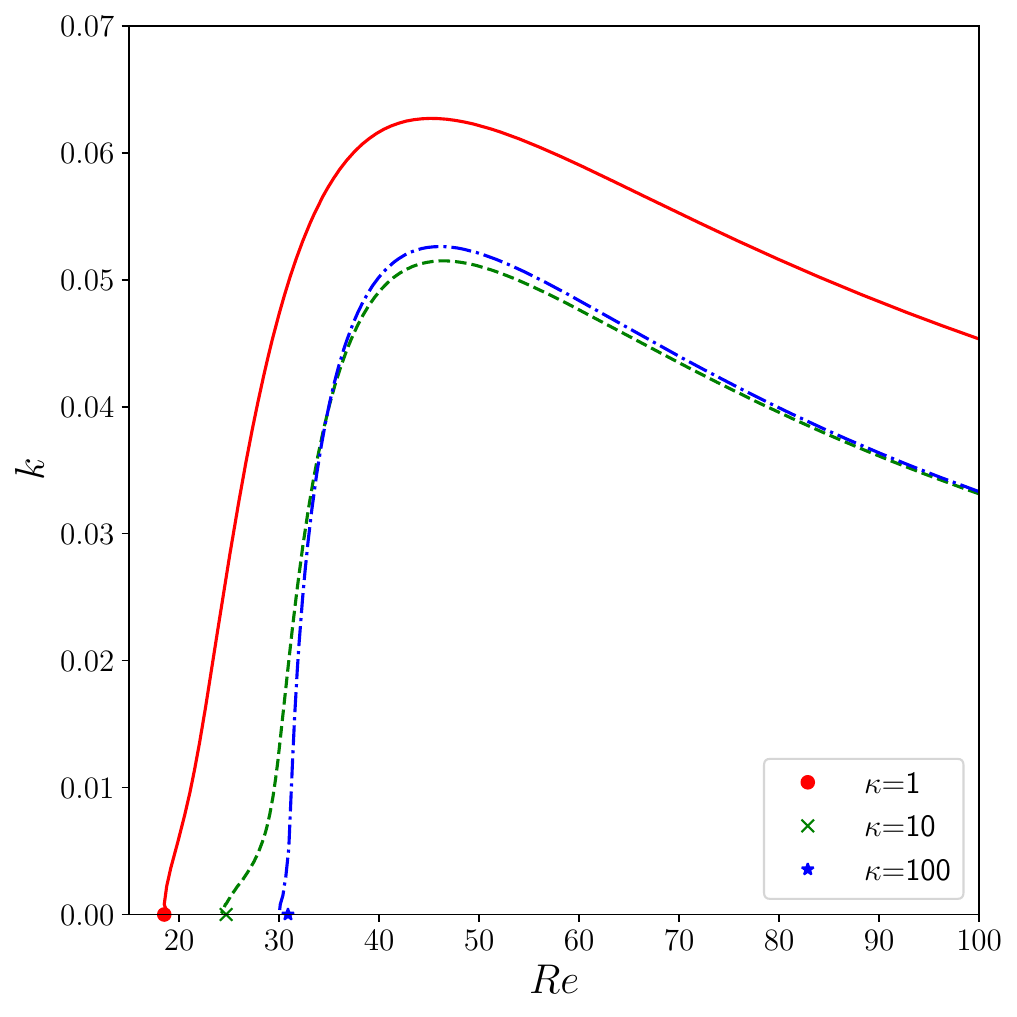}}%
		\hfill 
		\subcaptionbox{$Re = 40$}
		{\includegraphics[width=0.49\textwidth]{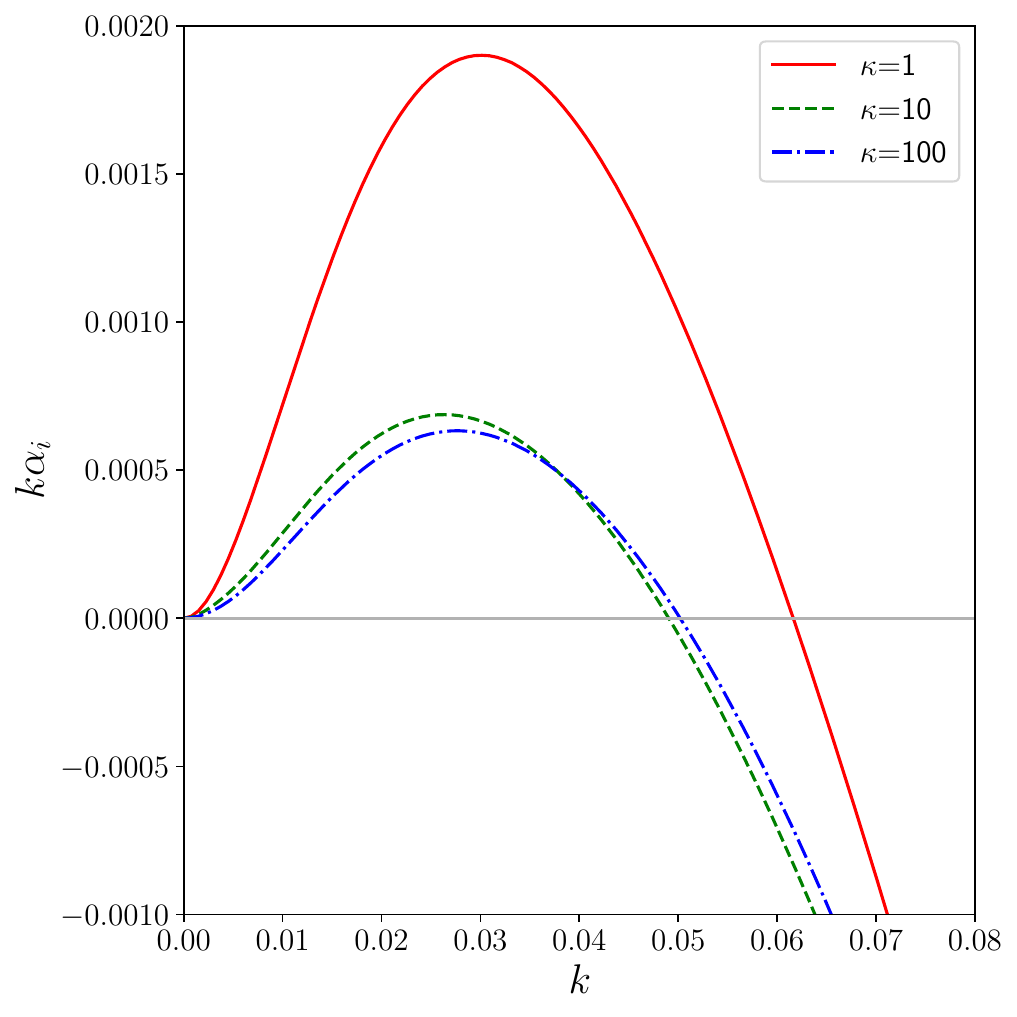}}
		\hfill 
		\caption{Effect of $\kappa$ in linear stability. $Mn=0.5, \beta=0.04, \Gamma_e=0.5, Pe_b = Pe_s=700, k_s=0.5, \theta = 4^{\circ}, Ka=1000$. Left panel: marginal stability curve, right panel: growth rate curve for $Re = 40$.  The marker in panel (a) represents the critical Reynolds number obtained through asymptotic analysis, but the other legends remain consistent.}
		\label{fig:liear_influence_kappa}
	\end{figure}    
	
	\begin{figure} 
		\subcaptionbox{}
		{\includegraphics[width=0.49\textwidth]{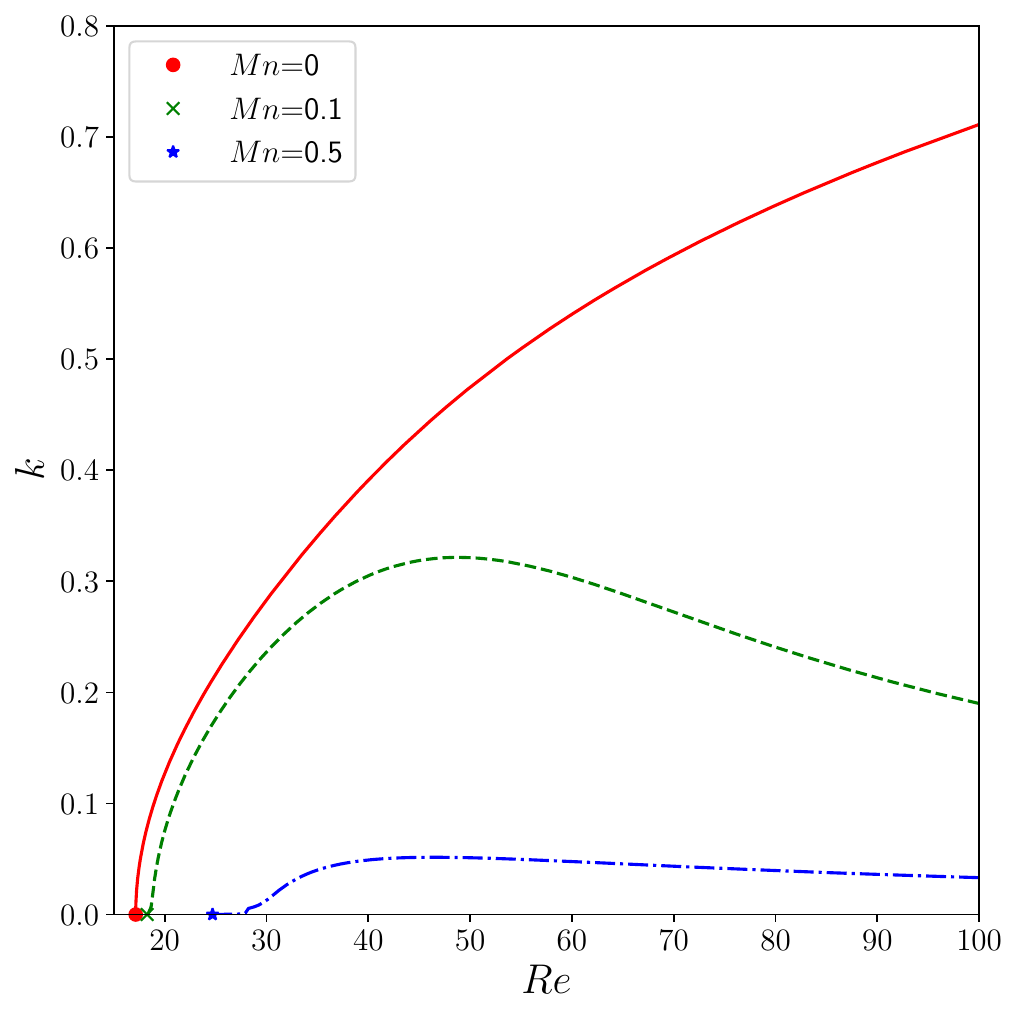}}%
		\hfill 
		\subcaptionbox{$Re = 40$}
		{\includegraphics[width=0.49\textwidth]{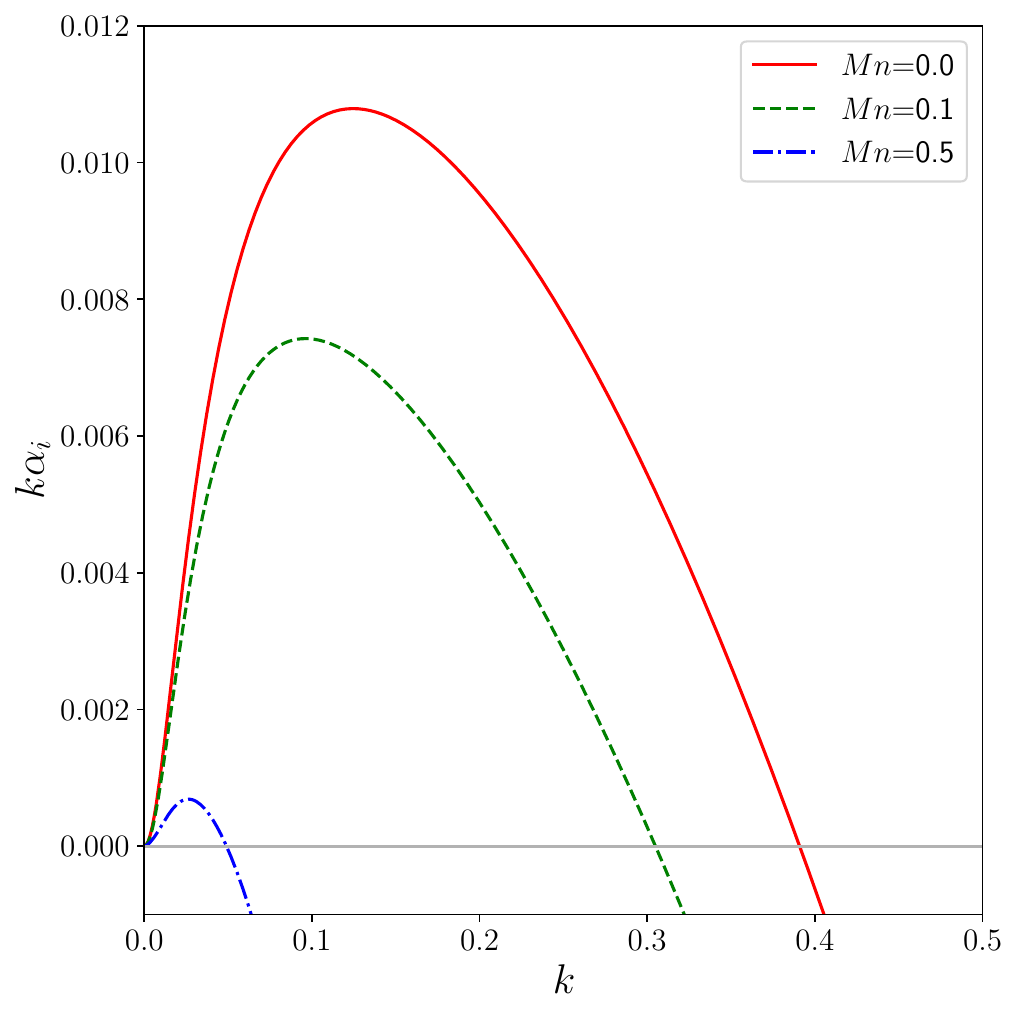}}
		\hfill 
		\caption{Effect of Marangoni number ($Mn$) in linear stability. $\beta=0.04, \Gamma_e=0.5, \kappa = 10, Pe_b = Pe_s=700, k_s=0.5, \theta = 4^{\circ}, Ka=1000$. Left panel: marginal stability curve, right panel: growth rate curve for $Re = 40$.  The marker in panel (a) represents the critical Reynolds number obtained through asymptotic analysis, but the other legends remain consistent.}
		\label{fig:liear_influence_Mn}
	\end{figure} 
	\par
	
	Figures \ref{fig:liear_influence_beta}–\ref{fig:liear_influence_Mn} show the neutral stability and temporal growth rate curves in the $Re-k$ plane obtained from numerical solutions to the Orr-Sommerfeld eigenvalue problem in different parametric spaces, for all wavenumbers (or for an arbitrary Reynolds number). Neutral stability curves separate the stable and unstable regions, and the interpretation of these curves at $k=0$ yields the critical Reynolds number for the amplification of infinitely long perturbations. Figures \ref{fig:liear_influence_beta} illustrate that the wall slip parameter $\beta$ has a dual effect on the primary instability. It is clear from the figures that near the onset of instability (small $Re$), increasing $\beta$ leads to increased growth rates and a higher peak in the dispersion curve at small $k$ (long waves); this implies that the slip promotes long-wave instability near the instability threshold.  Physically, slip at the substrate reduces shear resistance, allowing surface disturbances to grow more easily. This behavior of the wall-slip parameter aligns closely with the analytical findings based on the long-wave assumption near the instability threshold. In contrast, far from the critical
	Reynolds number we observe that beyond a certain critical value of the wavenumber $k_{+}$ (say), increasing $\beta$ causes a reduction in growth rate and then slip dampens the disturbances after that critical wavenumber ($k > k_{+}$). The wavenumber $k_{+}$ represents the intersection of two curves and is determined by the system parameters.  We observed that far from the threshold, $k_{+}$ shifts to the left as $Re$ increases. This phenomenon can be interpreted as the smoothing effect of slip on relatively high-frequency perturbations; possibly, for a fixed Reynolds number, the thickness of the base film decreases to compensate for the effect of the slip length, thus strengthening the impact of the capillary term on the primary instability. Therefore, increasing the wall-slip parameter causes a destabilizing effect near the onset of instability and plays a dual role further from the threshold; it initially destabilizes and then stabilizes after a critical wavenumber is reached. Numerous authors, including \citet{bhatLinearStabilityContaminated2018} and \citet{davalos-orozcoThermocapillaryStabilityViscoelastic2024}, have reported on the dual role of the slip parameter.
	Figures \ref{fig:liear_influence_Gamma_e} show the fluctuation of the marginal stability and growth rate curves with the equilibrium concentration of surface surfactant $\Gamma_e$. It is also evident from the figures that the growth rate curves intersect at a specific value of $k$, which plays a dual function in the primary instability as anticipated in the asymptotic analysis of the Orr-Sommerfeld study, which has been extensively studied previously.
	Figures \ref{fig:liear_influence_kappa} show that equilibrium constant adsorption $\kappa$ exerts a stabilizing effect as the growth rate declines with the persistent increase in $\kappa$. This observation also supports our earlier findings regarding the long-wave approximation. Figures \ref{fig:liear_influence_Mn} show that the Marangoni surfactant number has a stabilizing effect in the whole range of $Re$ unconditionally. This result is also in excellent agreement with the analytical findings for the long-wave assumption.
	
	\section{ Depth-averaged (DA) model}
	This section investigates how nonlinearity affects flow stability. One way to make things easier is to project the dynamics of the film onto a limited set of variables, or degrees of freedom, that are functions of the location $x$ on the plane and of time $t$ only. We use the \citet{ruyer-quilImprovedModelingFlows2000} weighted residual technique to create a four-equation model that is valid up to a moderate Reynolds number, consistent in first order, and includes second-order streamwise viscous terms.
	Since the normal velocity component of the plane $w = - \bigint_ 0^z \dfrac{\partial u}{\partial x} dz$ is relatively smaller than the stream component $u$, we can neglect the inertia terms and the viscous terms of the stream in the $z$ component of the momentum equation (\ref{y-momentum}). The inertia and the viscous term in the streamwise $z$ component of the momentum equation are of higher order and can therefore be omitted \citep{samantaFallingFilmSlippery2011}. The pressure distribution is obtained by integrating the remaining equation up to the order $\varepsilon$. Substituting the value of $p$ into equation (\ref{x-momentum}) produces an approximate momentum equation.
	
	\begin{eqnarray} \label{reduced_mom}
		\varepsilon Re \left(\frac{\partial u}{\partial t} + u \frac{\partial u}{\partial x} + w \frac{\partial u}{\partial z}\right) = \frac{Re}{Fr^2} + \frac{\partial^2 u}{\partial z^2} 
		- \varepsilon \frac{Re}{Fr^2}\cot\theta h_x 
		\nonumber\\
		+ 2 \varepsilon^2 \frac{\partial^2 u}{\partial x^2} + \varepsilon^2 \frac{\partial}{\partial x}\left[\frac{\partial u}{\partial x}\biggm|_{z = h}\right]
		+ \varepsilon Ka h_{xxx},
	\end{eqnarray}
	where $Ka =  {\varepsilon}^2 Re We$. Let us decompose the velocity components as: $u= u_0 + \varepsilon u_1+ \varepsilon^2 u_2 + ...,\quad w = \varepsilon w_1+ \varepsilon^2 w_2 + ...$. Then equation (\ref{reduced_mom}) reduces to
	\begin{eqnarray} \label{perturbed_mom}
		\frac{\partial^2 (u_0 + \varepsilon u_1)}{\partial z^2} + \frac{Re}{Fr^2} = \varepsilon Re \left(\frac{\partial u_0}{\partial t} + u_0 \frac{\partial u_0}{\partial x} + \varepsilon w_1 \frac{\partial u_0}{\partial z}\right) + \mathcal{K} 
		\nonumber\\
		+ \varepsilon \frac{Re}{Fr^2}\cot\theta \frac{\partial h}{\partial x} 
		- \varepsilon Ka \frac{\partial^2 h}{\partial x^2} - 2 \varepsilon^2 \frac{\partial^2 u_0}{\partial x^2} - \varepsilon^2 \left[\frac{\partial u_0}{\partial x}\biggm|_h \right]  + O(\varepsilon^3) \quad
	\end{eqnarray}
	where $\mathcal{K}$ represents the second order inertia corrections contributed by the deviations $u_1$ from the parabolic velocity profile and $\varepsilon w_1 = - \bigints_0^z \dfrac{\partial u_0}{\partial x} dz$. Moreover, to single out $u_1$ we use the gauge condition $\bigints_0^h u_1 dz = 0$, which ensures that $q$ still corresponds to the local flow rate.\\
	The main objective of the weighted residual method is to eliminate the dependence on $z$ by making certain assumptions about the velocity and solute concentration. We suggest the following specific profile for $u$ and $C$: 
	\begin{eqnarray}
		u_0(x, z,t) = a_0(x,t)f_0(\Bar{z}) + b_0(x,t) f_1(\Bar{z})
	\end{eqnarray}
	where
	\begin{eqnarray}
		f_0(\Bar{z}) = \Bar{z} - \frac{\Bar{z}^2}{2} + \Bar{\beta} \label{f_0}
		\\
		f_1(\Bar{z}) = 2\Bar{z} -  3 \Bar{z}^2\label{f_1}
	\end{eqnarray}
	with $\Bar{\beta} = \beta/h$ and $\Bar{z} = z/h$, and,
	\begin{eqnarray}
		a_0(x,t) = \dfrac{q(x,t) - \bigints_0^h u_1 dz}{h \bigints_0^1 f_0(\Bar{z}) d\Bar{z}}
		\\
		b_0(x,t) = Ma \dfrac{\partial \Gamma}{\partial x} \left(\dfrac{h}{4}\right)
	\end{eqnarray}
	
	\begin{eqnarray}
		C = \phi(x,t) + Pe_b k_s  \left(\frac{h}{2}\right)[\kappa (1 - \Gamma)\phi - \Gamma] (1 - \Bar{z}^2)
	\end{eqnarray}
	where $q(x,t) = \int_0^h u dz$, $Ma = \varepsilon Re Mn$ and  
	$\phi(x,t)  = \begin{cases}
		C(x, z = h, t) & \text{if }\quad t > 0 \\
		\phi_e = C_e & \text{if} \quad t = 0
	\end{cases}$
	
	The excess / deficit amount of surfactant in the bulk is given by $$\chi = \begin{cases} \int_0^h (C - \phi)dz = \dfrac{Pe_b k_s}{3} h^2 [\kappa (1 - \Gamma)\phi - \Gamma] & \text{if}\quad t > 0 \\ (C_e- \phi_e)h_e= 0 & \text{if}\quad t=0, \end{cases}$$ which is related to the net surfactant adsorption rate. Equation (\ref{perturbed_mom}) explains the balance of gravity acceleration and drag forces, which is influenced by inertia, viscous stream-wise diffusion, surface tension and the hydrostatic pressure gradient. The weighted residual approach renders the residual orthogonal to each element of a complete set; we may choose weight functions to emphasize speeded up by judicious choice of the weight functions. According to the Galerkin approach, we have appropriately chosen $f_0$ as the weight function and multiplied the equation (\ref{perturbed_mom}) by $f_0$ and integrated with respect to $z$ from $0$ to $h(x, t)$ to obtain a second-order weighted residual model for the momentum equations. We chose unity as the weight function for the concentration equation (\ref{concentration}) and integrated it from $z = 0$ to $z = h(x,t)$. To accurately preserve the critical condition, the momentum equation must be consistent up to order $\varepsilon$ in the averaging process, as the onset of the instability is captured by the gradient expansion up to first order \citep{ruyer-quilImprovedModelingFlows2000}. \par
	The weighted residual technique provides a convenient shortcut to obtain the evaluation equation for $q$ by carefully choosing the weight function $\mathtt{w}$. Writing equation (\ref{perturbed_mom}) as $BL(u_1)=0$, we introduce the inner product $\langle a, b \rangle = \int_0^h ab\;\mathrm{d}z$ derived from the plane norm $\textit{L}^2$ and the residual $\mathcal{R} =  \langle BL(u_1), \mathtt{w} \rangle = 0$ with the mass balance equation forms a system of evolution equation. Similarly, let us introduce the linear drag operator $\mathcal{L} = \partial_{zz}$ the only term involving $u_1$ in the residual $\mathcal{R}$ comes from the calculation of the drag $\langle \mathtt{w}, \mathcal{L} u_1 \rangle = \langle \mathcal{L}^* \mathtt{w}, u_1 \rangle$, where $\mathcal{L}^*$ denotes the adjoint operator of $\mathcal{L}$. Two integrations by parts suffice to show that $\mathcal{L}= \mathcal{L}^*$, that is, the operator $\mathcal{L}$ is self-adjoint. The order of $\varepsilon$ contribution of the drag can be canceled using the gauge condition $\langle u_1, 1 \rangle = 0$.
	
	After some simple algebraic calculation \citep{samantaFallingFilmSlippery2011, dalessioMarangoniInstabilitiesAssociated2020}, the model for variables $h$, $q$, $\chi$ and $\Gamma$ is as follows:
	\begin{eqnarray}\label{model_h}
		\frac{\partial h}{\partial t} + \frac{\partial q}{\partial x} = 0
	\end{eqnarray}
	\begin{eqnarray}\label{model_q}
		\dfrac{\partial q}{\partial t} + F(\bar{\beta}) \dfrac{q}{h} \dfrac{\partial q}{\partial x} -  G(\bar{\beta}) \dfrac{q^2}{h^2} \dfrac{\partial h}{\partial x} =  \dfrac{1}{\varepsilon Re}H(\bar{\beta}) \left[\left \lbrace \dfrac{Re}{Fr^2} \left(\dfrac{1 + 3 \bar{\beta}}{3}\right) h-\dfrac{q}{h^2} \right\rbrace  \right.
		\nonumber\\
		\left. - Ma \left(\dfrac{2 + 5\bar{\beta}}{4}\right)\dfrac{\partial \Gamma}{\partial x}\right]+ \dfrac{1}{Re} I(\bar{\beta})\left[\text{Ka} \: h \dfrac{\partial^3 h}{\partial x^3} - \dfrac{Re}{Fr^2} \cot\theta h \frac{\partial h}{\partial x}\right]
		\nonumber\\
		+ \frac{\varepsilon}{Re} \left[ J(\bar{\beta})\dfrac{q}{h^2} \left(\dfrac{\partial h}{\partial x}\right)^2 - K(\bar{\beta})\frac{1}{h} \dfrac{\partial q}{\partial x} \dfrac{\partial h}{\partial x} - L(\bar{\beta}) \dfrac{q}{h} \dfrac{\partial^2 h}{\partial x^2} + M(\bar{\beta}) \dfrac{\partial^2 q}{\partial x^2}\right]
		\nonumber\\
		+ Ma \left[P(\bar{\beta})h^2 \dfrac{\partial^2 \Gamma}{\partial t\partial x} + R(\bar{\beta}) q h \dfrac{\partial^2 \Gamma}{\partial x^2} + S(\bar{\beta}) h \dfrac{\partial q}{\partial x}\dfrac{\partial \Gamma}{\partial x} + T(\bar{\beta}) q \dfrac{\partial h}{\partial x}\dfrac{\partial \Gamma}{\partial x} \right]
	\end{eqnarray}
	\begin{eqnarray}\label{model_chi}
		\frac{\partial}{\partial t}
		\left( \chi + \phi h \right)
		+ \frac{\partial}{\partial x} \left[
		\biggl\{\left( \frac{33}{40} + 3 \bar{\beta} \right)\bigg/ (1 + 3\bar{\beta})\biggr\}  \frac{q\,\chi}{h}
		+ \phi q
		\right]
		\nonumber\\
		=
		\frac{\varepsilon}{\mathrm{Pe}_b} \left[
		\frac{\partial^2 \chi}{\partial x^2}
		- \frac{3\chi}{h^2} \left( \frac{\partial h}{\partial x} \right)^2
		+ h \frac{\partial^2 \phi}{\partial x^2}
		\right] 
		\nonumber\\ 
		- \frac{3}{\varepsilon Pe_b} \frac{\chi}{h^2} 
		- \frac{\mathrm{3 Ma}}{80}
		\left[
		\frac{\partial^2 \Gamma}{\partial x^2} h \chi
		+ \frac{\partial \chi}{\partial x} \frac{\partial \Gamma}{\partial x} h
		+ \chi \frac{\partial \Gamma}{\partial x} \frac{\partial h}{\partial x}
		\right]
	\end{eqnarray}
	\begin{eqnarray}\label{model_Gamma}
		\frac{\partial \Gamma}{\partial t} + \frac{\partial}{\partial x} \biggl\{\frac{3(1 + 2 \bar{\beta})}{2(1 + 3 \bar{\beta})} \left(\frac{q \Gamma}{h}\right)\biggr\} = \frac{\varepsilon}{Pe_s} \frac{\partial^2 \Gamma}{\partial x^2} + \frac{3}{\varepsilon Pe_b}\frac{\chi}{h^2}
		\nonumber\\
		+ \frac{1}{4} Ma \left[\frac{\partial^2 \Gamma}{\partial x^2} \Gamma h + \left(\frac{\partial \Gamma}{\partial x}\right)^2 h + \Gamma\frac{\partial \Gamma}{\partial x}\dfrac{\partial h}{\partial x} \right] 
	\end{eqnarray}
	
	where,
	\begin{eqnarray}
		F(\bar{\beta}) &=& \dfrac{34 + 14\bar{\beta} \lbrace 17 + 45 \bar{\beta} (1 +\bar{\beta})\rbrace}{7  \lbrace 2 + 5\bar{\beta}(2 + 3 \bar{\beta})\rbrace (1 + 3 \bar{\beta})}
		\\
		G(\bar{\beta}) &=& \dfrac{18 + 9\bar{\beta} [18 + 35\bar{\beta}\lbrace 2 + \bar{\beta}(4 + 3 \bar{\beta})\rbrace]}{7  \lbrace 2 + 5\bar{\beta}(2 + 3 \bar{\beta})\rbrace (1 + 3 \bar{\beta})^2}
		\\
		H(\bar{\beta}) &=& \dfrac{5 (1 + 3 \bar{\beta})}{ \lbrace 2 + 5\bar{\beta}(2 + 3 \bar{\beta})\rbrace}
		\\
		I(\bar{\beta}) &=& \dfrac{5 (1 + 3 \bar{\beta})^2}{3  \lbrace 2 + 5\bar{\beta}(2 + 3 \bar{\beta})\rbrace}
		\\
		J(\bar{\beta}) &=& \dfrac{8 + 3 \bar{\beta} [24 + \bar{\beta}\lbrace 107 + 45 \bar{\beta} (5 + 4 \bar{\beta})\rbrace]}{ \lbrace 2 + 5\bar{\beta}(2 + 3 \bar{\beta})\rbrace (1 + 3 \bar{\beta})^2}
		\\
		K(\bar{\beta}) &=& \dfrac{9 (1 + 2 \bar{\beta})\lbrace 1 + 5\bar{\beta}(1 + 2\bar{\beta})\rbrace}{ \lbrace 2 + 5\bar{\beta}(2 + 3 \bar{\beta})\rbrace (1 + 3 \bar{\beta})}
		\\
		L(\bar{\beta}) &=& \dfrac{6 (1 + 2 \bar{\beta})}{(1 + 3 \bar{\beta})}
		\\
		M(\bar{\beta}) &=& \dfrac{3 \lbrace 3 + 5 \bar{\beta}(3 + 4 \bar{\beta})\rbrace}{ \lbrace 2 + 5\bar{\beta}(2 + 3 \bar{\beta})\rbrace}
		\\	P(\bar{\beta}) &=& \dfrac{(1 + 3 \bar{\beta})}{24  \lbrace 2 + 5\bar{\beta}(2 + 3 \bar{\beta})\rbrace}
		\\
		R(\bar{\beta}) &=& \dfrac{3(5 + 14\bar{\beta})}{112  \lbrace 2 + 5\bar{\beta}(2 + 3 \bar{\beta})\rbrace}
		\\
		S(\bar{\beta}) &=& \dfrac{19 + 42 \bar{\beta}}{168  \lbrace 2 + 5\bar{\beta}(2 + 3 \bar{\beta})\rbrace}
		\\
		T(\bar{\beta}) &=& \dfrac{5 + 12 \bar{\beta}(4 + 7 \bar{\beta})}{56  \lbrace 2 + 5\bar{\beta}(2 + 3 \bar{\beta})\rbrace (1 + 3 \bar{\beta})}
	\end{eqnarray}

	The model equations (\ref{model_h} - \ref{model_chi}) are identical to those derived by \citet{dalessioMarangoniInstabilitiesAssociated2020} for isothermal films with soluble surfactants in the no-slip limit $\beta = 0$, but equation (\ref{model_Gamma}) is slightly different in the multiplier (1/4) instead of (5/4) in the last term. This discrepancy arises due to the additional term in equation (\ref{boundary_surface_sarfactant}), as pointed out earlier. This disparity causes a mass increase in the free surface as time progresses in the nonlinear simulation, impeding the system's ability to reach a stable solution over time.\footnote{For more details, please refer to Appendix C.} Note that the source term in the surface equation (\ref{model_Gamma}) $+ \frac{3}{\varepsilon Pe_b} \frac{\chi}{h^2}$ appears in the bulk equation (\ref{model_chi}) as $- \frac{3}{\varepsilon Pe_b} \frac{\chi}{h^2}$ with the same coefficient $\frac{3}{\varepsilon Pe_b}$ with the opposite sign, as we have defined $\chi$ appropriately.  
	Equations (\ref{model_h} - \ref{model_Gamma}) are identical with \cite{pereiraDynamicsFallingFilm2008}   (equations (45a-45c) of their paper; equation (\ref{model_chi}) will be redundant) in the insoluble limit, with the velocity scaling $\hat{V} = 2 \hat{U}_N$, twice the dimensional Nusselt free surface velocity, and with $\beta =0$, for no-slip substrate.  It is worth noting that  \citet{pereiraDynamicsFallingFilm2008} calculated the equations up to the order of $\varepsilon$, but this is correct up to the order of $\varepsilon^2$.
	Equations (\ref{model_h} - \ref{model_q}) reduce to the simplified second-order momentum equation derived by \citet{samantaFallingFilmSlippery2011} in the case of zero surfactant, and finally to that derived by \citet{ruyer-quilImprovedModelingFlows2000} in the no-slip limit $\beta = 0$. The two remaining equations (\ref{model_chi}) and (\ref{model_Gamma}) of the model will be superfluous for the no-surfactant limit.

	\subsection{Numerical Solution of DA model}
	Equations (\ref{model_h})-(\ref{model_Gamma}) are numerically solved with a pseudospectral discretization coupled with a high-order time integration. 
	Due to periodic boundary conditions, each unknown $f$ is decomposed as a truncated Fourier series:
	\begin{equation}
		f(x,t) = \sum_{k=-N/2}^{N/2} \hat{f_k}(t)~e^{ikx}
	\end{equation}
	with $N$ the number of collocation points \citep{Peyret::book}.
	
	Spatial derivatives are then computed using the following:
	\begin{equation}
		\frac{\partial^m}{\partial x^m} f(x,t) = \sum_{k=-N/2}^{N/2}(ik)^m \hat{f_k}(t)~e^{ikx}
	\end{equation}
	The aliasing behavior is mitigated through the de-aliasing process,\textit{ i.e.}
	by filtering the modes above $k_f = 2k_{max}/(p+1)$, with $p$ the order of nonlinearity \citep{orszag1971elimination} and $k_{max}=(\pi/L)N$. Temporal derivatives are
	integrated with a fourth-order Runge-Kutta scheme \citep{carpenter1994rk,Peyret::book}.
	Finally, the mixed term $\displaystyle{\dfrac{\partial^2 \Gamma}{\partial t\partial x}}$
	of equation (\ref{model_q}) is evaluated by calculating the spatial derivative
	of equation (\ref{model_Gamma}) :
	\begin{eqnarray}
		\dfrac{\partial^2 \Gamma}{\partial t\partial x} = \dfrac{\partial}{\partial x}\dfrac{\partial\Gamma}{\partial t} =
		\dfrac{\partial}{\partial x} \left [ \frac{\varepsilon}{Pe_s} \frac{\partial^2 \Gamma}{\partial x^2} +  \frac{3}{\varepsilon Pe_b}
		\frac{\chi}{h^2} - \frac{\partial}{\partial x} \biggl\{\frac{3(1 + 2 \bar{\beta})}{2(1 + 3 \bar{\beta})} \left(\frac{q \Gamma}{h}\right)\biggr\} \right ] \nonumber \\ 
		+ \dfrac{\partial}{\partial x} \left [\frac{1}{4}Ma \left\{ \frac{\partial^2 \Gamma}{\partial x^2} \Gamma + \left(\frac{\partial \Gamma}{\partial x}\right)^2 h + \Gamma\frac{\partial \Gamma}{\partial x}\dfrac{\partial h}{\partial x}  \right \} \right ].
	\end{eqnarray}
	
	Finally, we monitor the total surfactant mass
	\begin{equation}
		M_{\mathrm{tot}}(t)
		=
		\int_0^L \left( \Gamma(x,t) + \int_0^{h(x,t)} C(x,z,t)\,dz \right) dx,
	\end{equation}
	and verify that its relative variation remains at the level of numerical round-off
	for all simulations reported below.
	This diagnostic is essential here, since earlier long-wave models
	\citep{pascalStabilityInclinedFlow2019,dalessioMarangoniInstabilitiesAssociated2020}
	exhibit spurious interfacial mass growth in long-time simulations
	(see Appendix~C).

	\begin{figure}
		\centering
		\includegraphics[width=0.5\textwidth]{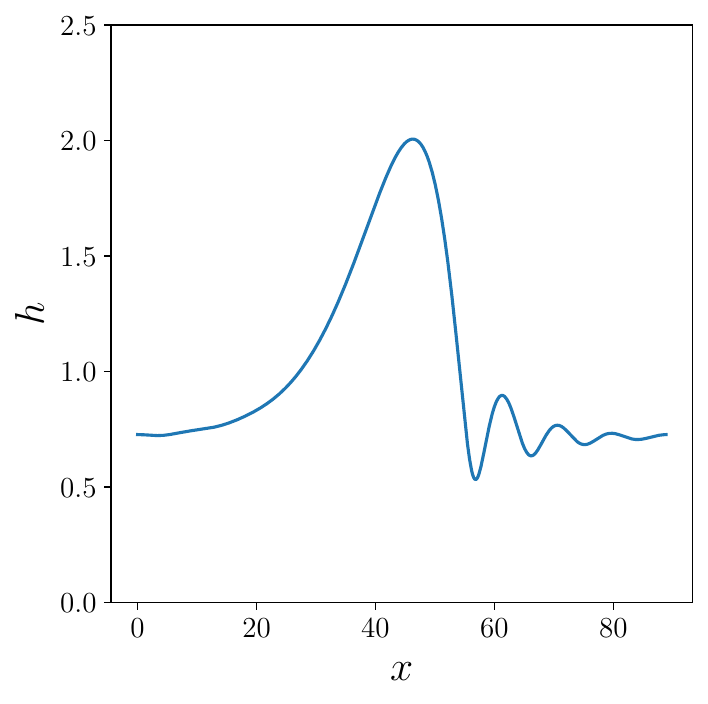}
		\caption{\label{fig:samanta_fig_12} Validation of our program by reproducing Figure 12 of \cite{samantaFallingFilmSlippery2011} for the surfactant-free case.}
	\end{figure}

	\begin{figure}
		\centering
		\includegraphics[width=\textwidth]{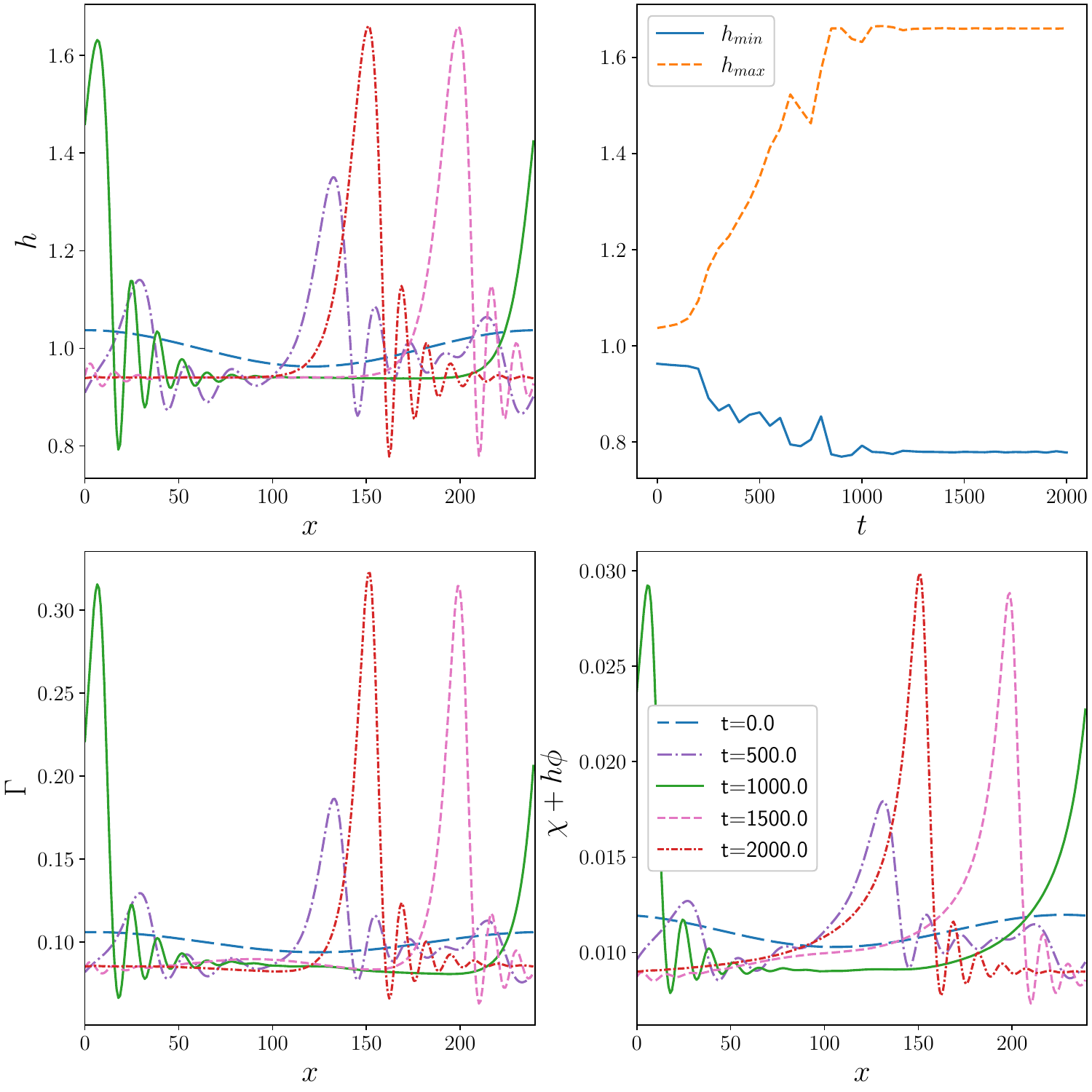}
		\caption{\label{fig:base_figure_non_linear} Formation of solitary waves over the time for $\beta = 0.04$, $\Gamma_e = 0.1$, $Mn = 0.1$ and $\kappa = 10$ $\delta = (Re-Re_c)/Re_c \simeq 1.9$, indicating a strongly supercritical regime.}
	\end{figure}
	
	In falling-film flows, the wave shape and amplitude depend strongly on how far the system
	is from the instability threshold.
	To distinguish genuine parameter effects from a trivial change in supercriticality, we therefore quantify the distance from the onset using
	\begin{equation}
		\delta = \frac{Re - Re_c}{Re_c},
	\end{equation}
	where $Re_c$ is evaluated for the corresponding parameter set. When parameters such as $\beta$ or $\Gamma_e$ are varied at fixed $Re$, the corresponding change in $Re_c$ implies that the system moves across different supercriticality levels. Reporting $\delta$ is therefore essential to distinguish genuine parameter-induced changes in wave structure from trivial changes caused by varying distance from onset.
	
	We verified numerical convergence by increasing the number of Fourier modes and reducing the time step until
	changes in wave amplitude and phase speed were below a prescribed tolerance. We also monitor the total surfactant mass $M_{\mathrm{tot}}$ and confirm that its relative variation remains within numerical round-off throughout the simulations. To benchmark and validate our program, we reproduced Figure 12(a) of \cite{samantaFallingFilmSlippery2011} as a limiting case without surfactant and obtained identical results  (Figure \ref{fig:samanta_fig_12}), thus confirming the precision of our implementation.\par
	
	The figures in this section are plotted for $Re = 50,$ $Ka = 1000,$ $\theta = 4^\circ,$ $k_s = 0.5,$ $Pe_s = Pe_b = 700,$ if not mentioned otherwise.

	\begin{figure}
		\centering
		\includegraphics[width=1\textwidth]{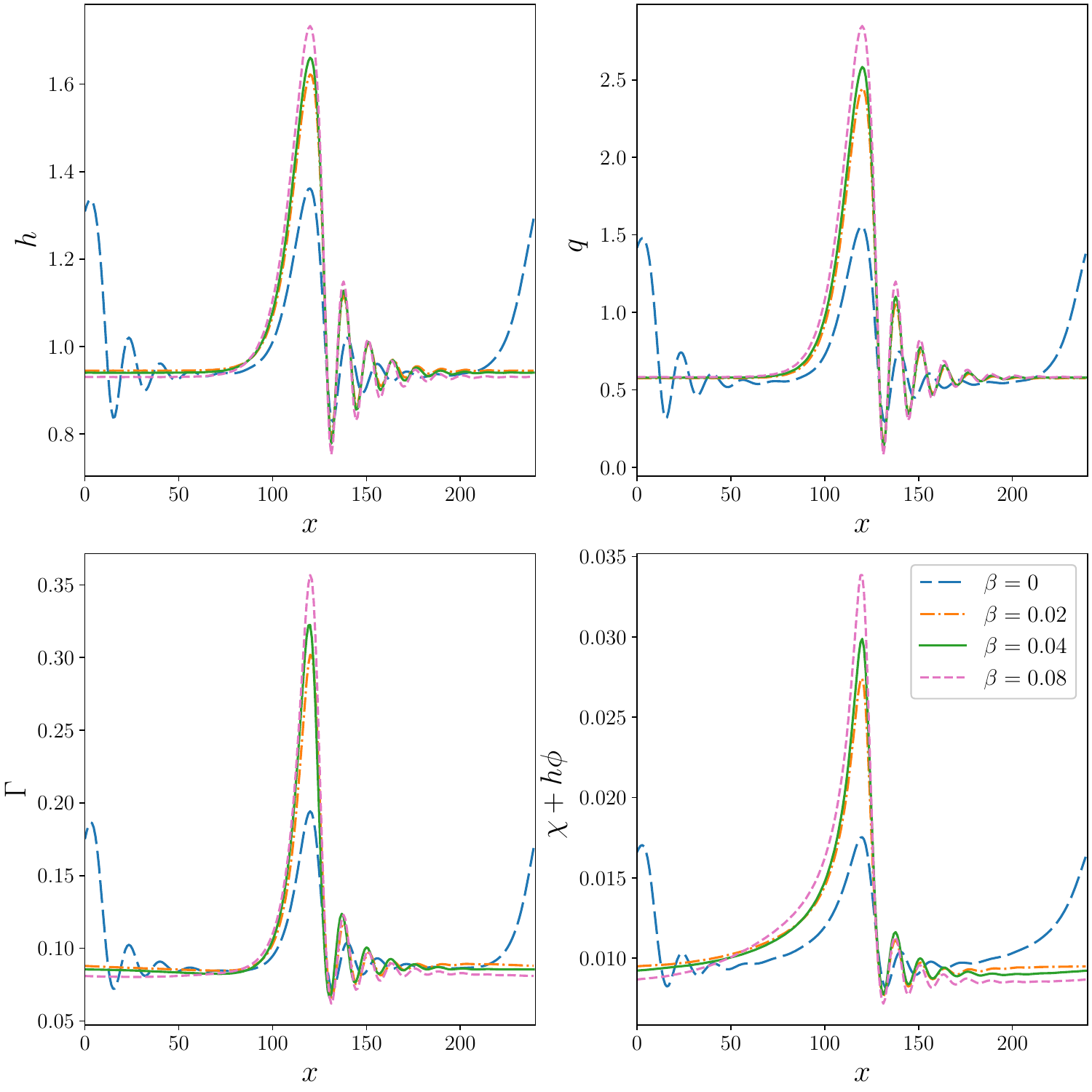}
		\caption{\label{fig:beta_var_figure_non_linear} Permanent solitary waves at time $t= 2000$ for different slip variable $\beta$, when $\Gamma_e = 0.1$, $Mn = 0.1$ and $\kappa = 10$,}
	\end{figure}
	
	\begin{figure}
		\centering
		\includegraphics[width=1\textwidth]{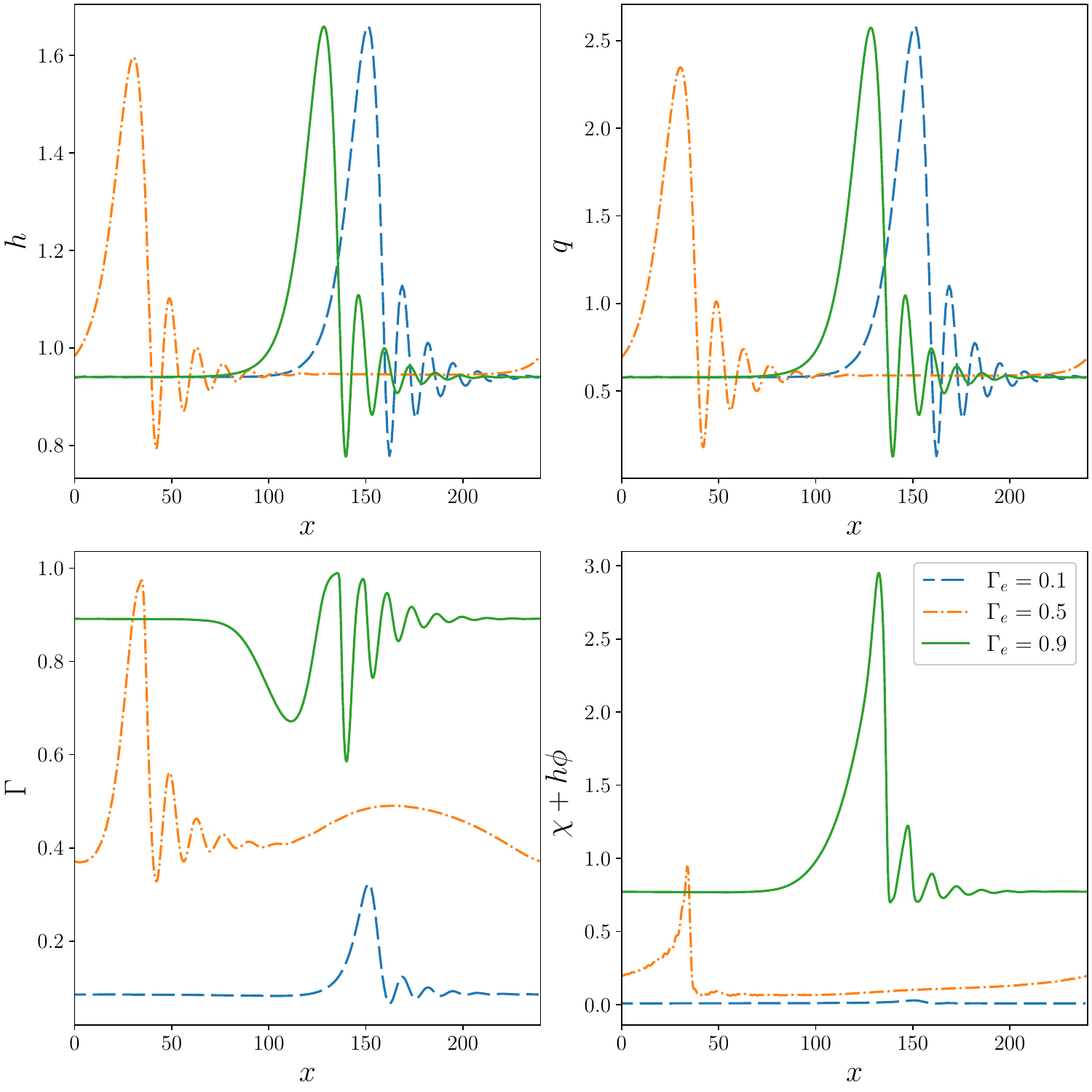}
		\caption{\label{fig:gammaE_var_figure_non_linear} Permanent solitary waves at time $t= 2000$ for different $\Gamma_e$, when $\beta = 0.04$, $Mn = 0.1$ and $\kappa = 10$}
	\end{figure}

	\begin{figure}
		\centering
		\includegraphics[width=1\textwidth]{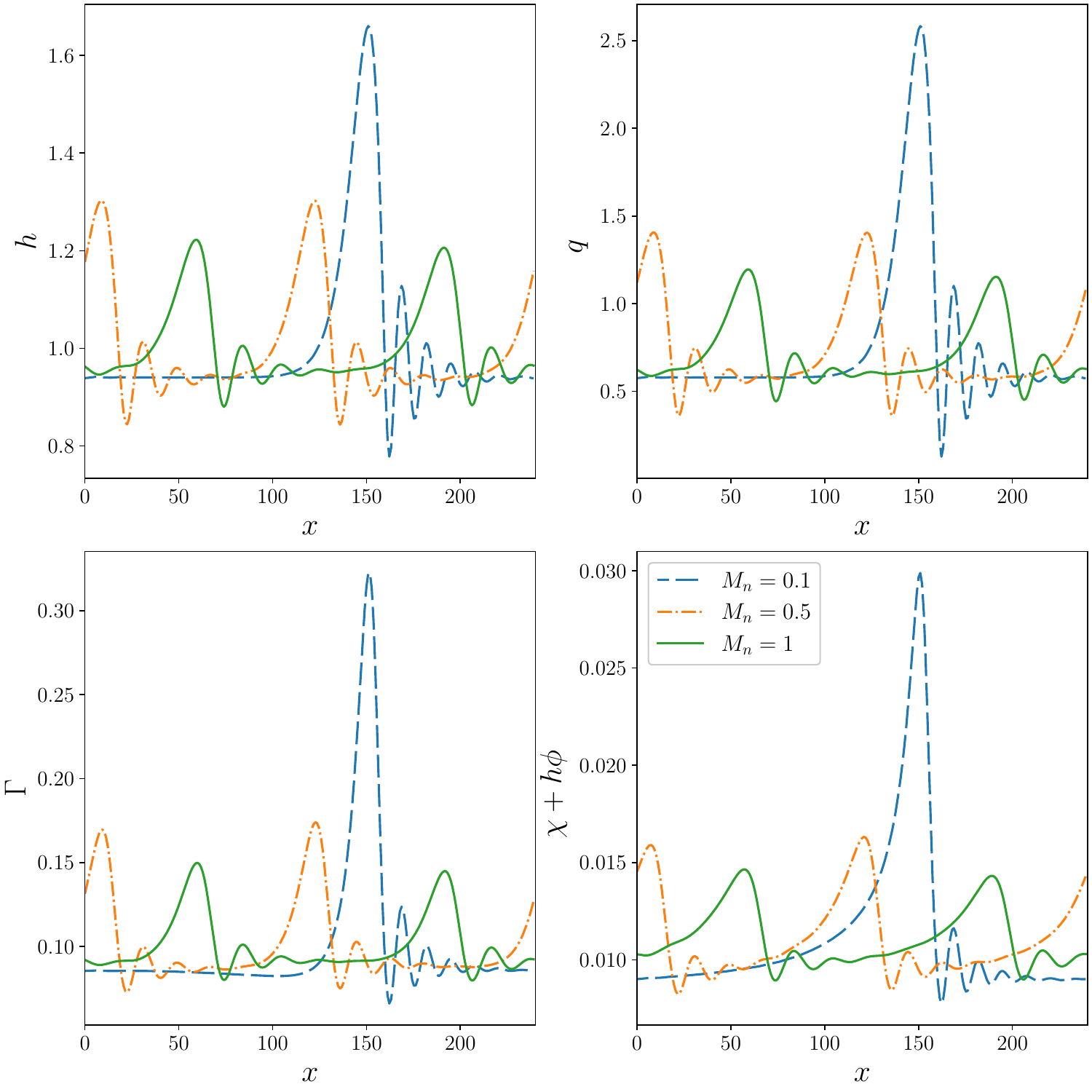}
		\caption{\label{fig:Mn_var_figure_non_linear} Permanent solitary waves at time $t= 2000$ for different $Mn$, when $\beta = 0.04$, $\Gamma_e = 0.1$ and $\kappa = 10$}
	\end{figure}
	
	\begin{figure}
		\centering
		\includegraphics[width=1\textwidth]{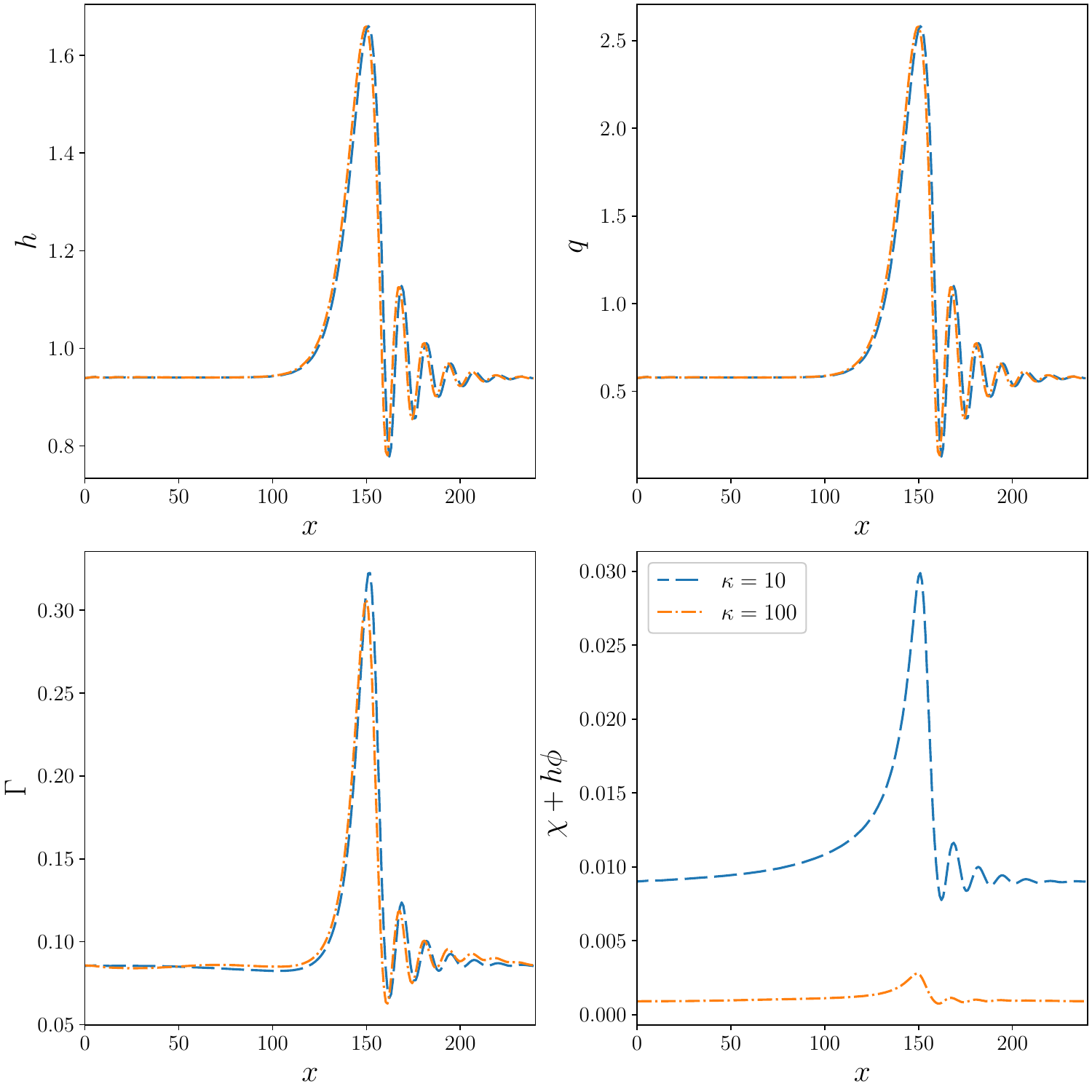}
		\caption{\label{fig:kappa_var_figure_non_linear} Permanent solitary waves at time $t= 2000$  for different $\kappa$, when $\beta = 0.04$, $\Gamma_e = 0.1$ and $Mn = 0.1$}
	\end{figure}

	Figure \ref{fig:base_figure_non_linear} tracks the time evolution of the film height $h(x,t)$, the interfacial surfactant $\Gamma(x,t)$, and the bulk inventory $\chi(x,t)+h(x,t)\phi(x,t)$ in a periodic domain of length $L=240$, starting from a small disturbance aligned with the leading eigenvector of the linearized system (\ref{model_h})–(\ref{model_Gamma}). The disturbance rapidly organizes into nonlinear permanent waves. The envelopes of $h_{\min}(t)$ and $h_{\max}(t)$ level off to a quasi-steady plateau by $t\approx1200$ (for the parameter set in the caption), while $\Gamma$ and $\chi{+}h\phi$ settle into smoothly translating profiles. This behavior contrasts with \citet{pascalStabilityInclinedFlow2019} and \citet{dalessioMarangoniInstabilitiesAssociated2020}, who reported an unstable evolution of $q$ and $\Gamma$ and a time-independent $\chi$. In our formulation, a fully conservative bulk–surface balance eliminates the spurious gain/loss of interfacial mass, so $\chi$ evolves consistently with advection–diffusion in the bulk and adsorption at the interface, and the coupled system selects stable traveling waves rather than drifting or ill-posed states. This rectified mass budget is essential for observing saturation of wave amplitudes and for capturing the long-time permanence of the nonlinear structures shown.
	\par

	Figures \ref{fig:beta_var_figure_non_linear},  \ref{fig:gammaE_var_figure_non_linear}, \ref{fig:Mn_var_figure_non_linear} and  \ref{fig:kappa_var_figure_non_linear}  show how film thickness $h$, flow rate $q$, surface surfactant concentration $\Gamma$, and total surfactant content in the bulk $\chi+ h \phi$ change along the streamwise direction on fully developed permanent waves (at t = 2000) as parameters $\beta$, $\Gamma_e$, $Mn$ and $\kappa$ vary. \par

	All of the profiles in Figure \ref{fig:beta_var_figure_non_linear} have been phase-aligned, with the main crest near the center of the computational domain. Under no-slip conditions ($\beta = 0$), the interface forms a two-hump permanent wave. Wall slip $(\beta \neq 0)$ reduces wall friction, permitting liquid to spread easier. The pattern settles to a single hump, with enhanced capillary ripples ahead of the main wave hump.  Slip minimizes near-wall shear and Marangoni back-stress (which favors single-crest selection while suppressing the bound pair), but the greater local flux $q$ associated with slip feeds the short-wave capillary branch, sharpening the ripple field.
	
	The surface concentration $\Gamma$ and the bulk content  $\chi + h\phi$ exhibit matching corrugations, tracking the interfacial shape and advective transport. To achieve thickness control and defect suppression in coating, lubrication, and microfluidic applications, increasing $\beta$  reduces multi-crest formations to a single crest but results in stronger capillary ripples. \par
	
	Figure \ref{fig:gammaE_var_figure_non_linear}  shows the variations for three distinct equilibrium surface surfactant concentrations ( $\Gamma_e = 0.1, 0.5$ and $0.9$ ) for $\beta = 0.04$, $Mn = 0.1$, $\kappa = 10$ and $Re = 50$. The interface shows a non-monotonic tendency, with $h$ (similar to $q$) nearly the same for $\Gamma_e = 0.1$ and $0.9$, but becoming slightly smaller when $\Gamma_e = 0.5$.  
	
	When the surface is almost clean ($\Gamma_e = 0.1$), interfacial tension dominates and the film remains relatively stable; when it is heavily contaminated ($\Gamma_e = 0.9$), surface elasticity weakens because the interface is close to saturation. At the intermediate value, i.e., around $\Gamma_e \approx 0.5$, even small perturbations in $\Gamma$ maximize surface tension gradient, creating strong Marangoni stresses that oppose interfacial deformation and slightly reduce the film height. The effect is most apparent in $\Gamma$, which exhibits its largest spatial fluctuations at $\Gamma_e \approx 0.5$. This intermediate state corresponds to the maximum in $Re_c$ observed earlier in linear analysis (Figure \ref{fig:linear_analytic}). \par
	
	From Figure \ref{fig:Mn_var_figure_non_linear} it is evident that at low $Mn = 0.1$, the film exhibits a single broad hump preceded by a pronounced train of capillary ripples. As $Mn$ increases to $0.5$ and $1$, Marangoni stresses strengthen and the pattern reorganizes: the main elevation splits into two solitary-like crests, while the capillary ripples are strongly suppressed. Physically, the enhanced interfacial elasticity at larger $Mn$ generates surface-compression/extension zones that redistribute momentum beneath the interface (supporting two peaks) and, at the same time, damp the short-wave capillary branch, erasing the high-frequency oscillations prominent at low $Mn$.\par
	
	To demonstrate the variation of the adsorption equilibrium constant $\kappa$ in Figure \ref{fig:kappa_var_figure_non_linear}, we have chosen two values $\kappa = 10$ and $100$ while keeping the rest the same as in Figure \ref{fig:gammaE_var_figure_non_linear} with $Mn = 0.1$. 
	The parameter $\kappa$ characterizes the strength of exchange between the bulk and the interface: higher values correspond to faster adsorption–desorption kinetics and a more equilibrated surface. As seen in the figure, increasing $\kappa$ produces only minor changes in the film height and flow rate, indicating that the overall hydrodynamic field remains largely unaffected by the kinetic exchange in this parameter regime. However, the surfactant-related fields respond more sensitively. With $\Gamma_e$ fixed, increasing $\kappa$ mainly affects the bulk field: $\chi + h \phi$ loses most of its modulation and approaches an almost smooth, nearly flat profile. By contrast, $\Gamma$ keeps essentially the same baseline shape, with only minor amplitude/phase adjustments. This reflects faster adsorption–desorption, rapidly re-equilibrating the bulk with the interface. It plays a key role in homogenizing the interfacial surfactant distribution and stabilizing the film against concentration-driven disturbances.

	\section{Conclusion}
	In this work we have examined the stability and nonlinear dynamics of a gravity-driven thin liquid film laden with a
	soluble surfactant flowing down an inclined substrate with wall slip.
	Slip at the solid boundary was modeled using a Navier condition, while surfactant transport was described through
	a coupled bulk-surface formulation with adsorption-desorption kinetics.
	The linear stability problem was analyzed by solving the full Orr-Sommerfeld eigenvalue system, both asymptotically
	in the long-wave limit and numerically for arbitrary wavenumbers.
	
	The linear analysis highlights a fundamental difference between soluble and insoluble surfactants.
	While insoluble surfactants are known to have a purely stabilizing influence, soluble surfactants exhibit a dual
	role: the critical Reynolds number initially increases with the equilibrium surface coverage $\Gamma_e$, indicating
	stabilization, but decreases beyond a threshold value, leading to destabilization.
	This non-monotonic behavior reflects the competing effects of Marangoni stresses and bulk-surface exchange.
	Both the Marangoni number and the solubility parameter tend to damp interfacial disturbances and delay the onset of
	instability.
	
	The numerical solution of the full eigenvalue problem confirms these asymptotic predictions and clarifies the role
	of wall slip.
	Close to the instability threshold, slip enhances the growth of unstable modes, whereas at higher Reynolds numbers
	it has a dual effect: long waves are destabilized, while shorter waves are stabilized.
	As the Reynolds number increases, the boundary between these regimes shifts toward smaller wavenumbers, resulting
	in a widening of the stable band.
	
	To investigate the post-critical dynamics, a reduced long-wave model was derived using a weighted-residual approach.
	The model consistently incorporates slip, Marangoni stresses, and soluble surfactant transport, and correctly
	recovers the limiting cases of clean films, insoluble surfactants, and no-slip substrates.
	Nonlinear simulations reveal that increasing the Marangoni number drives a transition from single-hump to
	double-hump solitary-wave structures, accompanied by a reduction in capillary ripples and a non-monotonic
	dependence on the equilibrium surface coverage.
	
	An important outcome of the present formulation is the resolution of the nonphysical accumulation of surfactant
	mass reported in earlier long-wave studies \citep{pascalStabilityInclinedFlow2019,dalessioMarangoniInstabilitiesAssociated2020}.
	By enforcing strict global conservation of total surfactant mass, the model yields physically meaningful steady
	interfacial states and long-time dynamics.
	
	Overall, the results show that wall slip, quantified by the parameter $\beta$, provides a practical control
	mechanism for surfactant-laden falling films.
	By systematically linking $\beta$ to instability onset, wave selection, ripple attenuation, and the bulk-surface
	surfactant balance, we demonstrate how slip suppresses fragile multi-hump bound states and promotes either a single,
	broad crest or an almost flat, uniform film.
	The underlying mechanism is straightforward: slip reduces near-wall shear and Marangoni back-stress, weakens
	tail-tail interactions between waves, and promotes homogenization of the interfacial surfactant field.
	From an applied perspective, this suggests that surface chemistry or microstructural design of substrates can be
	used to improve film uniformity and robustness.
	Finally, we emphasize that the slip condition is employed here as an effective boundary condition; in genuinely
	porous or textured substrates, the slip length may depend on microstructural details and flow conditions, which are
	not resolved in the present model.
	
	The present framework provides a consistent basis for future studies incorporating thermal effects, non-Newtonian rheology, or dynamically structured substrates, where the interplay between interfacial transport and wall slip is expected to be even richer.

	\subsection*{Acknowledgment}
	The authors would like to thank the French Institute in India (IFI) - the Embassy of France in India for providing Scientific High-Level Visiting Fellowships (SSHN) 2024 to initiate this project, and heartfelt gratitude to Professor Christian Ruyer-Quil for insightful discussions and advice. 
	
	\subsection*{Declaration of Interests}
	The authors report no conflict of interest.
	
	\subsection*{Appendix A}
	\subsubsection*{Surfactant transport equation at the free surface}
	Equation (\ref{surface_surfactant}) is written in vector form in terms of the surface gradient operator $\nabla_s$ to maintain consistency with previous works in the literature, but it is appropriate to write the equation in tensor form in curvilinear coordinates since the free surface $z = h(x,t)$ is a curve in the $xz$ plane, and equation (\ref{surface_surfactant}) is valid on the free surface. Then, equation (\ref{surface_surfactant}) in tensorial form is given below:
	\begin{equation}\label{tensor_surfactant}
		\frac{\partial \Gamma}{\partial t} + \text{div} (\Gamma V_i) = D_s\text{div}(\text{grad}\Gamma)
		+ J_{bs},
	\end{equation}
	where $V_i = u(x, z=h(x,t), t)$ are the components of the velocity on the free surface. 
	The components of velocity in curvilinear coordinates are indeed a covariant tensor, though velocity itself is a contravariant vector. Therefore, we have to consider the divergence of a co-/contravariant vector in terms of the covariant derivative (since partial derivatives are not a tensor). To find the explicit form of equation (\ref{tensor_surfactant}), we have to find the divergence of a covariant vector and the Laplacian of a scalar function in a curvilinear coordinate system. For, let $\mu^i$ be an arbitrary contravariant vector, and let $\mu_i$ be the corresponding covariant form. Then their covariant derivatives are $\mu^i_{,j}$ and $\mu_{i,j}$, respectively. Then their divergence is given by
	$$\text{div}  \mu^i = \mu^i_{,i} = \frac{1}{\sqrt{g}} \frac{\partial}{\partial x^l}(\sqrt{g} \mu^l) = \text{div}  \mu_i = \mu_{i,i} =  \frac{1}{\sqrt{g}} \frac{\partial}{\partial x^l}(\sqrt{g} g^{lj}\mu_j),$$ where, $g^{ij}$ is the reciprocal tensor of the fundamental tensor $g_{ij}$ and $\mu^i= g^{ij}\mu_j$ and $g = \text{det}g_{ij} \neq 0$. \\
	Note that in this case the Laplacian of a scalar function $\phi$ in a curvilinear coordinate system is the divergence of this gradient vector field is $$\Delta \phi = \nabla^2 \phi = \text{div}(\text{grad} \phi) = \frac{1}{\sqrt{g}} \frac{\partial}{\partial x^i} \left( \sqrt{g} g^{ij} \frac{\partial \phi}{\partial x^j} \right).$$  where, $g^{ij}$ is the reciprocal tensor of the fundamental tensor $g_{ij}$ and $\mu^i= g^{ij}\mu_j$.
	
	Thus, the explicit form of equation(\ref{tensor_surfactant}) is 
	\begin{equation}\label{tensor_surfactant_explicit}
		\frac{\partial \Gamma}{\partial t} + \frac{1}{\sqrt{g}} \frac{\partial}{\partial x^l}(\sqrt{g} g^{lj}(\Gamma V_j)) =  D_s \frac{1}{\sqrt{g}} \frac{\partial}{\partial x^i} \left( \sqrt{g} g^{ij} \frac{\partial \Gamma}{\partial x^j} \right)
		+ J_{bs},
	\end{equation}
	where,  $g_{11} = (1+h^2_x),\quad$ $g^{11} = 1/(1+h^2_x)$ along the stream direction, since
	$ds^2 = dx^2+dz^2 = (1+h^2_x)dx^2 = g_{11}dx^2$. It is not surprising that normalization is not needed when writing equation (\ref{tensor_surfactant_explicit}), since the tensor equation is independent of coordinate transformation.
	\subsection*{Appendix B}
	\subsubsection*{The interface mode and the surfactant mode}
	The interface and surfactant modes can be derived analytically from the small wavenumber expansion of the Orr-Sommerfeld eigenvalue problem, by suitably normalizing $\psi_0$ and deleting replicated solutions of $\psi_0$ that occur at each of the higher orders, which are as follows:  
	\begin{eqnarray}
		\alpha^{(1)} = (1 + 2 \beta)\dfrac{Re}{Fr^2} + i k Re \left[\dfrac{(1 + \beta)\lbrace2 + 5\beta(2 + 3\beta)\rbrace}{15}\left(\dfrac{Re}{Fr^2}\right)^2 \right.
		\nonumber\\
		\left. - \dfrac{(1 + 3 \beta)}{3}\left(\dfrac{Re}{Fr^2}\right) \dfrac{\cot\theta}{Re} - \dfrac{3 (1 + 2 \beta)Mn \Gamma_e \kappa(1-\Gamma_e)^2}{4 + 3 \kappa(1 - \Gamma_e)^2} \right] +O(k^2)   
	\end{eqnarray}
	\begin{equation}
		\alpha^{(2)} = (1 + 2 \beta) \dfrac{1}{2}\dfrac{Re}{Fr^2} - \dfrac{ik}{Pe_s} + O(k^2)
	\end{equation}
	
	\begin{figure} 
		\centering
		\subcaptionbox{}
		{\includegraphics[width=0.5\textwidth]{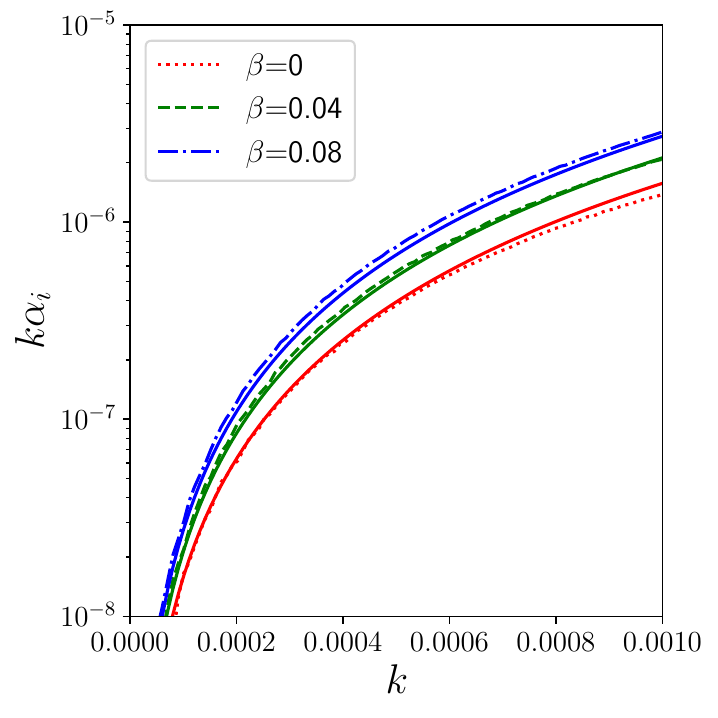}}%
		\hfill 
		\subcaptionbox{}
		{\includegraphics[width=0.5\textwidth]{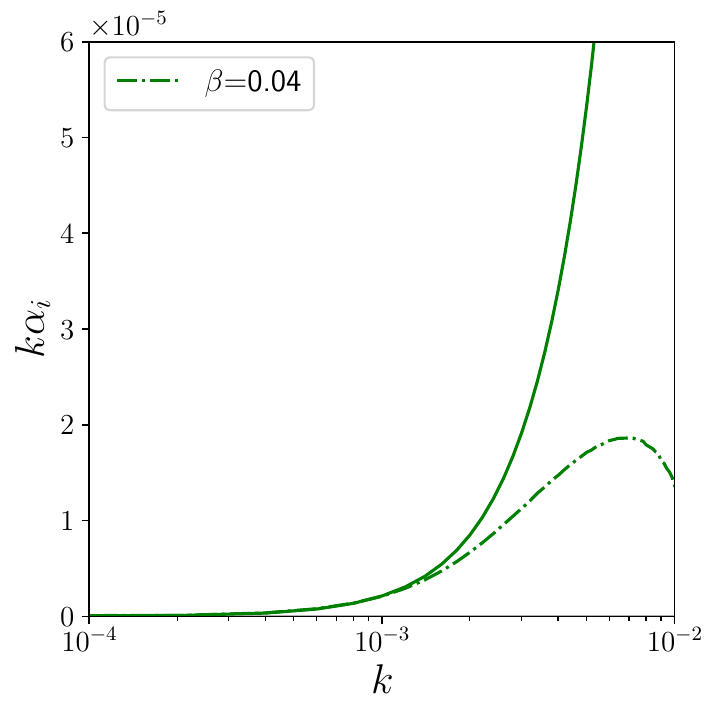}}
		\hfill 
		\caption{Comparison of growth rate with that from analytic (solid lines) and numerical (dashed lines) results of Orr Sommerfeld analysis for $Re = 30$: (a) $k \alpha_i$ - axis is in log-scale; (b) $k$-axis log scale plotted only for $\beta = 0.04$.}
		\label{fig:growth_kapitza_mode}
	\end{figure}

	The above two modes are quite similar to their insoluble surfactant counterparts $\kappa \to \infty$, as revealed by \cite{pereiraDynamicsFallingFilm2008} in the non-slip limit $(\beta = 0)$.  The interface mode is activated by interface deflection, whereas the surfactant mode occurs when the amplitude of the surface surfactant changes, regardless of interfacial deflection. \\
	For the interface mode \citep{liRoleSolubleSurfactant2023}, we set $\eta_0=1$, $\eta_1=\eta_2=0$, and other variables change according to it. Wave celerity $\alpha_0$ is independent of surfactant solubility, which confirms the insoluble case. However, bulk surfactant advection and surfactant solubility affect $\gamma_0$ values.  \cite{karapetsasRoleSurfactantsMechanism2014}, as well as \cite{kalogirouRoleSolubleSurfactants2019}, reported similar findings. Similarly, for the surfactant mode \citep{liRoleSolubleSurfactant2023}, we set $\gamma_0 = 1$, $\gamma_1=\gamma_2=0$ and $\eta_0 = 0$ therefore $c_0= 1/\kappa(1-\Gamma_e)^2 $. \\
	In the non-slip limit, the third mode (\cite{pereiraDynamicsFallingFilm2008}), namely the shear mode, can be found as: 
	\begin{equation}
		\alpha^{(3)} = - \dfrac{\pi^2}{4kRe}n^2 + O(1), \quad n \quad \text{is odd}. 
	\end{equation}
	
	That is, countably infinite numbers of eigenvalues (\cite{pereiraDynamicsFallingFilm2008}). The classical shear mode loses its mathematical foundation under Navier slip conditions in the long-wave limit, as slip undermines the wall vorticity anchoring essential to its validity. The eigenvalues $\alpha^{(i)}$'s, representing the different modes of instabilities, are the unique solutions of the Orr-Sommerfeld problem in the limit of a small wavenumber \citep{pereiraDynamicsFallingFilm2008}. \cite{pereiraDynamicsFallingFilm2008} clarified that the surfactant mode is a diffusion mode modified by advection and associated with the surface transport equation. The surface mode is a classical long-wave interface instability mode, also known as the Kapitza mode \citep{kapitzaWaveFlowThin1949}, modified by Marangoni effects, while shear modes are Yih modes \citep{yihStabilityLiquidFlow1963} characterized by hydrodynamics (without Marangoni effects) associated with semiparabolic Nusselt velocity profiles.
	We would like to note that in the asymptotic solution of the OS problem section, we computed the onset of instability (critical Reynolds number) from the hydrodynamic Kapitza mode $\alpha^{(1)}$. \\
	
	The first mode defines the threshold for hydrodynamic instabilities and has been extensively discussed by various writers, beginning with Benjamin \cite{benjaminWaveFormationLaminar1957} and Yih \cite{yihStabilityLiquidFlow1963} for a clean fluid.  The phase velocity of this mode is $c_R=(1 + 2 \beta)\dfrac{Re}{Fr^2}+O(k^3)$ (the real part of $\alpha^{(1)}$, see equation (\ref{phase_speed})), which depends on the slip parameter and is twice the interfacial velocity of the flat film to the leading order in $k$.  In figure (\ref{fig:growth_kapitza_mode}), we compare the growth of the disturbance to the numerical solution of the Orr-Sommerfeld eigenvalue problem.  We observed that it is in good agreement for small wavenumbers and low to moderate Reynolds numbers.  The imaginary part of $\alpha^{(1)}$ denotes the onset of instability when the coefficient of $O(k^3)$ vanishes and the critical Reynolds number is given in (\ref{critical_soluble}) and is also discussed in detail in that section. \\
	
	The second mode, which we call the surfactant mode, is essentially a diffusional mode modified by advection and is related to the specific structure of the surface transport equation and is independent of the solubility of the surfactant, which is first detected by  \cite{blythEffectSurfactantStability2004} by solving the Orr-Sommerfeld problem.  \cite{pereiraDynamicsFallingFilm2008} identified the mode as a concentration mode and noted that it is connected with the diffusion and advection of species by flow, as evidenced by the presence of $Pe_s$. The phase velocity for this mode is $c_R = (1 + 2 \beta) \dfrac{1}{2}\dfrac{Re}{Fr^2}+O(k^3)$, which also depends on the slip parameter and is the same with the interfacial velocity of the flat film to the leading order in $k$, allowing the species to be transported by the flow.  This mode is strictly stable as long as the surfactant is able to diffuse; i.e., $Pe_s$ is finite. However, it becomes neutrally stable when $Pe_s\to \infty$ (\cite{blythEffectSurfactantStability2004}, \cite{pereiraDynamicsFallingFilm2008}). In which case convection is much faster than diffusion, and any perturbation in the concentration of surfactant does not decay but is advected by flow without change.

	\subsection*{Appendix C}
	\begin{figure}
		\centering
		\includegraphics[width=1.1\textwidth]{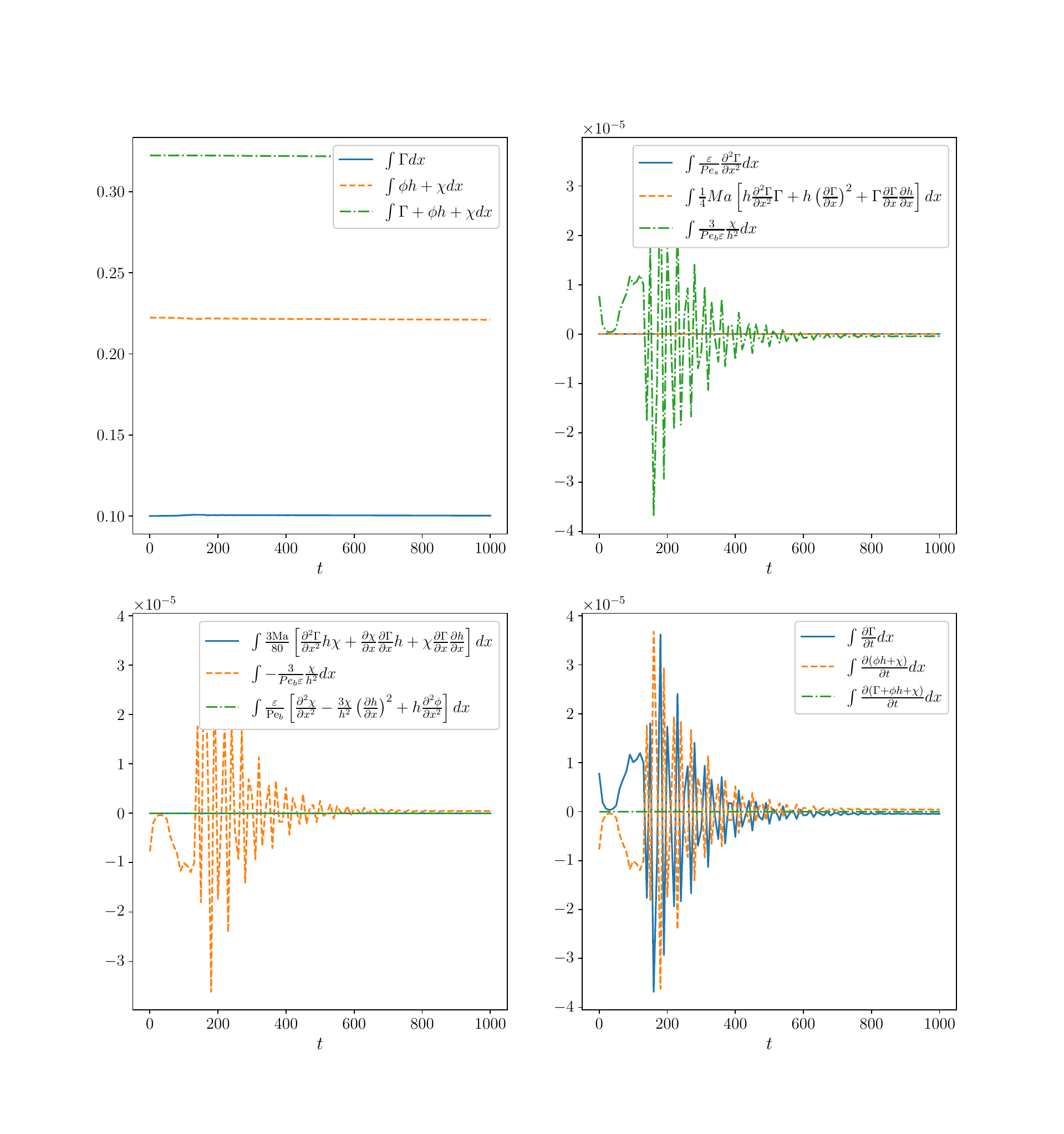}
		\caption{\label{fig::Mtot} Surface, bulk and total mass and mass flux for model equations (\ref{model_h} - \ref{model_Gamma}) with $\beta=0$. Upper left: Surfactant mass, bulk mass and total mass. Upper right: mass fluxes for equation (\ref{model_Gamma}). Lower left: mass fluxes for equation (\ref{model_chi}). Lower right: time derivative of surface, bulk and total mass. }
	\end{figure}
	\begin{figure}
		\centering
		\includegraphics[width=1.1\textwidth]{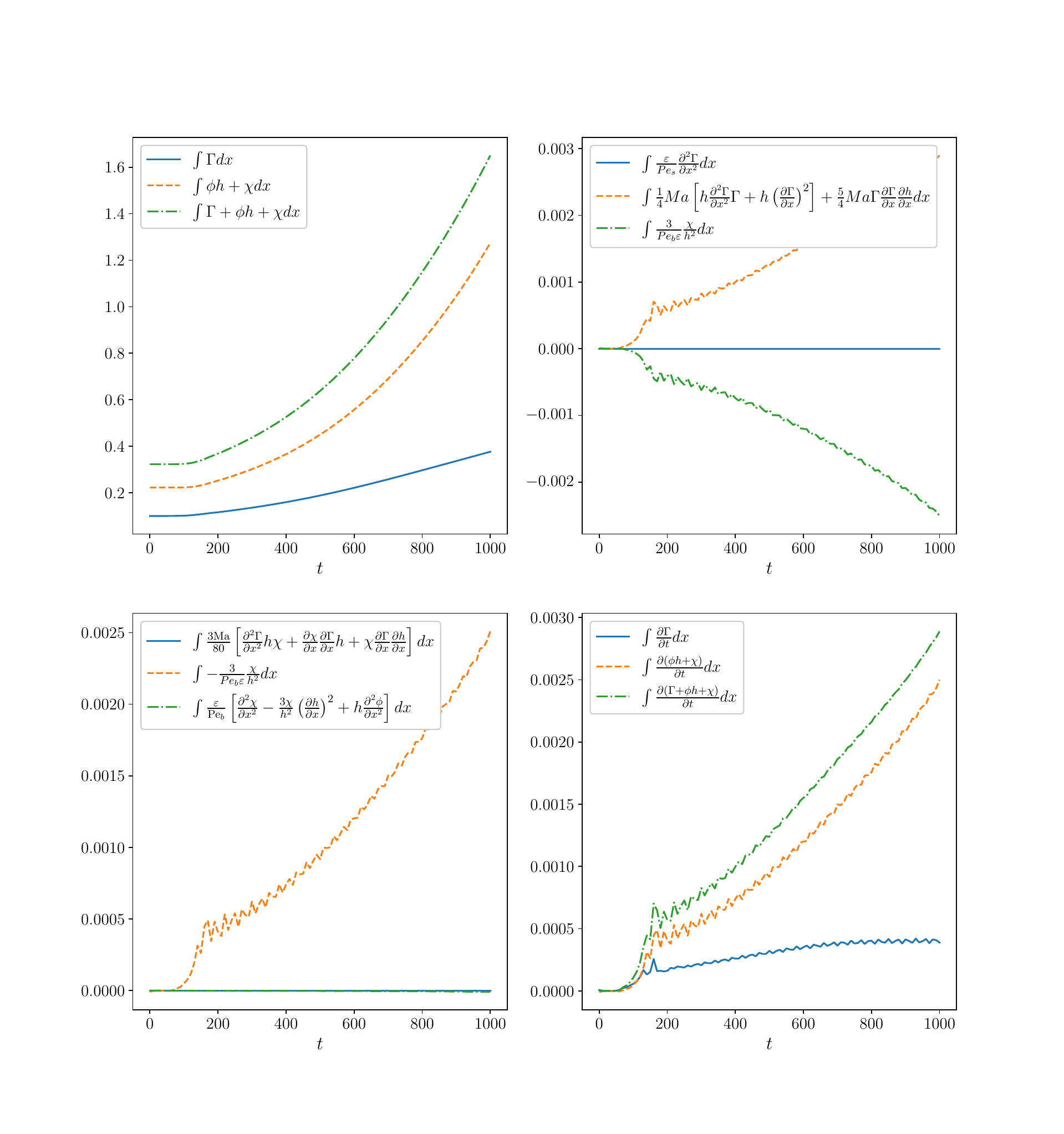}
		\caption{\label{fig::Mtot_wrong} Total mass and mass flux for model equations of \cite{pascalStabilityInclinedFlow2019} without density fluctuations. Upper left: Surfactant mass, bulk mass and total mass. Upper right: mass fluxes for surface concentration. Lower left: mass fluxes for bulk concentration. Lower right: time derivative of surface, bulk and total mass.}
	\end{figure}
	
	\subsubsection*{Surfactant mass conservation}
	The total mass of the surfactant at any instant $t$ is given by $$
	M(t) = \int_0^L \int_0^h C(x,z,t) dz dx + \int_0^L  \Gamma(x,t)  dx = \int_0^L [\phi(x,t) h (x,t) + \chi(x,t) + \Gamma(x,t) ] dx $$ is a constant quantity. 
	Since
	\[\frac{\partial M(t)}{\partial t} = 0 \implies 
	\frac{\partial}{\partial t}
	\int_{0}^{L}\Bigl[\phi(x,t)\,h(x,t)+\chi(x,t)+\Gamma(x,t)\Bigr]\;dx
	=0,
	\]
	it follows that
	\[
	\int_{0}^{L}\Bigl[
	\frac{\partial}{\partial t}(\chi + \phi\,h)
	\;+\;
	\frac{\partial \Gamma}{\partial t}
	\Bigr]\,dx
	=0.
	\]
	Since \(L\) is arbitrary, the integrand must vanish:
	\[
	\frac{\partial}{\partial t}(\chi + \phi\,h)
	\;+\;
	\frac{\partial \Gamma}{\partial t}
	=0.
	\]
	This conservation law provides a useful diagnostic for evaluating the consistency and fidelity of numerical models that involve soluble surfactants. \\
	
	Figure (\ref{fig::Mtot}) presents the temporal evolution of the surface and bulk surfactant masses, the total surfactant mass, and their respective fluxes. The surface mass increases slightly due to the interfacial transfer from the bulk, governed by the term $\frac{3}{\varepsilon Pe_b}\frac{\chi}{h^2}$. The corresponding decrease in the bulk mass preserves overall conservation, ensuring that the total surfactant mass remains constant. For comparison, figure (\ref{fig::Mtot_wrong}) shows the same quantities computed using the formulation of \cite{pascalStabilityInclinedFlow2019}  for a no-slip wall ($\beta = 0$) and in the absence of density variations. The only distinction between the two models is the numerical coefficient in the nonlinear term, which changes from $1/4 \rightarrow 5/4$ for the term $\Gamma \frac{\partial \Gamma}{\partial x} \frac{\partial h}{\partial x}$. This apparently minor modification produces an unnatural exponential growth of all surfactant mass components over time, revealing a fundamental inconsistency in the earlier closure.

	\bibliographystyle{jfm}
	\bibliography{bibliography}

\end{document}